\documentclass{ws-ijmpa}
\usepackage[super,compress]{cite}
\usepackage{graphicx}
\begin{document}
\markboth{Francesco Intravaia}{How modes shape Casimir Physics}

%
\catchline{}{}{}{}{}
%

\title{How modes shape Casimir Physics}

\author{Francesco Intravaia}
\address{Humboldt-Universität zu Berlin, Institut für Physik, 
\\ AG Theoretische Optik \& Photonik
12489 Berlin, Germany\\
francesco.intravaia@physik.hu-berlin.de
}

\maketitle


\begin{abstract}
Modes have steadily influenced the understanding of physical systems through time. At least since the prediction of the Casimir effect, they also play a very important role in Casimir Physics and in the understanding of the different phenomena typical of this research field. At equilibrium modes provide a direct connection between the zero-point energy and the existence of irreducible fluctuations in a quantum system, offering an anatomic view into the physics of the interaction. In nonequilibrium systems, modes can be decisive to understand the behavior of quantum fluctuation-induced phenomena, highlighting key aspects which determine their strength and their functional dependence. 
In this article we review some recent studies and results that highlight how modes impact Casimir physics and the central role they play in shaping our understanding of this area of research.
\end{abstract}

\renewcommand{\imath}{\mathrm{i}}


\tableofcontents

\section{Introduction}
\label{sec:1}

In a common definition, a mode is a stationary state of a system which oscillates in time without changing its relevant properties. Although, strictly speaking, this definition mostly corresponds to a mathematical abstraction, it reveals to be quite useful for understanding the physics of many systems. Indeed, their dynamics can be conveniently described as a combination of modes evolving according to their respective frequencies and physical features. As taking apart a complex engine and looking at each single gear helps to understand its functioning, the modal analysis provides a deep anatomic view into the physics of the system, possibly unveiling interesting and useful behaviors which will otherwise remain buried. One of the main advantages of this reductionist approach is that usually the analysis of a single mode is usually simpler than that of the whole system.

The advent of quantum mechanics, starting from the formulation of the Schr\"odinger equation to the more involved formalism of quantum field theory, has distinctively enhanced the interest in determining the modes of a system. Within these theories modes undergo various inflections, ranging from the definition of a quantum-mechanical eigenstate to the more general concept of eigenvector. They all have different physical meaning but a similar mathematical structure.
In Casimir physics the relevance of a mode decomposition permeates our understanding of the Casimir interaction until today. It elegantly interlaces with the definition of zero-point energy and dives deep into the physical and mathematical foundations of quantum theory of fluctuation-induced interactions. In both equilibrium and nonequilibrium systems, a mode analysis can provide a complementary point of view, which can help in understanding controversial results and allows for the prediction of a panoply of interesting and unintuitive phenomena. 

In this chapter we are going to illustrate with a few examples how modes shaped and still shape Casimir physics, connecting physical phenomena, theoretical results and functional behavior under a common denominator.

\section{Casimir's sum over modes approach}
\label{sec:2}

In 1948 H. B. G. Casimir~\cite{Casimir48} predicted the existence of a quantum force between two parallel, nonmagnetic and uncharged, perfectly reflecting parallel plates placed in vacuum. Following an inspiring discussion with Bohr~\cite{Milonni94}, his attention was directed to analyzing the system's electromagnetic zero-point energy. 
In electromagnetism, the combination of Maxwell equations and the boundary conditions specific to the system under analysis
allows for the existence of modes with a frequency dispersion relation $\omega_{\mathbf{K}}$, where $\mathbf{K}$ is usually a collections of parameters which uniquely characterize the mode. Formally, each mode behaves as a harmonic oscillator, which depending on the boundary conditions can be damped or not. As a consequence $\omega_{\mathbf{K}}$ is in general a complex-valued function. However, when every form of dissipation is suppressed, the frequencies are real and each mode contributes to the total ground state energy of the system with the value $\hbar\omega_{\mathbf{K}}/2$.
This is the case in Casimir's original calculation, where two different configurations are compared: First when the cavity formed by the plates has a finite length $L$ and then for $L\to \infty$. 
Mathematically this can be written using the following expression 
\begin{align}
E(L)&= \left[ \sum_{\mathbf{K}} \frac{\hbar\,\omega_{\mathbf{K}}}{2}\right]^{L}_{L\to\infty}, 
\label{sum-over-modes}
\end{align}
where, for simplicity, we have introduced the symbol $[\cdots]_{\infty}^{L}$, which indicates the difference between the total zero-point energies of the two configurations. For a cavity formed by two parallel planes one has  
\begin{equation}
\sum_{\mathbf{K}}\equiv\sum_{\sigma}A\int \frac{{\rm d^2}\mathbf{k}}{(2\pi)^{2}}\sum_{n}
\end{equation}
where $n$ is the index of the mode, $\sigma$ defines the polarization (TE or TM) of the corresponding electromagnetic field, $\mathbf{k}$ is the component of the wave vector parallel to the surfaces, and $A$ is the area of the planes. If the planes are  perfectly conducting, as assumed by Casimir, the mode frequency does not depend on $\sigma$ and is given by $\omega_{\mathbf{K}}=\omega^{\sigma}_{n}(\mathbf{k})=c\sqrt{|\mathbf{k}|^{2}+(n\pi/L)^{2}}$. 
Although each of the two configurations has an infinite zero-point energy their difference is finite, leading to the expression for the Casimir energy and force
\begin{equation}
E_{\rm Cas}(L)=-\frac{\hbar c}{720}\frac{A}{L^{3}}\qquad F_{\rm Cas}(L)=-\partial_{L}E_{\rm Cas}(L)=-\frac{\hbar c}{240}\frac{A}{L^{4}}~.
\label{EnergyForce}
\end{equation}

This result was so unexpected that even thirty years later Schwinger et al. called it
``\emph{one of the least intuitive consequences of quantum electrodynamics}'' \cite{Schwinger78}. 
There are many reasons for which the Casimir effect is so remarkable. Perhaps the most important one is that it associates a mechanical force with the vacuum, which in the classical view of physics was perceived as inert and absolute. Quantum mechanics dramatically changed this perspective, conferring to the vacuum some physical properties, thereby making it not unique and also observer-dependent.
In 1948 the zero-point energy was perceived as an unpleasant side-effect of quantum mechanics. Many scientists did not know what to do with it and some considered it as ``\emph{deprived of any physical reality}'' \cite{Pauli46}, something one should probably ignore. 
The typical arguments were that adding a constant to the Hamiltonian does not alter the physics of the system. Furthermore, energy is an integrated quantity and therefore only differences between energy values or with respect to a reference level have a physical meaning.

These or similar thoughts probably influenced Casimir's approach to the zero-point energy and led to the expression in Eq.~\eqref{sum-over-modes}, where the limit $L\to \infty$ is implicitly set as reference level with respect to which the energy of other configurations is measured. 
One of Casimir's most relevant insights was to recognize that the ground state energy of a quantum electromagnetic system is rather malleable: Despite being possibly infinite, it might substantially change its value. 
As long as this change is depending on a system parameter, the Casimir energy can at equilibrium be treated as a thermodynamic potential and the variation with respect to this parameter corresponds to the existence of the conjugated interaction. For instance, if the energy is distance dependent, as for Casimir's parallel plane setup, the conjugated interaction corresponds to a force, while we  have a torque if the energy is varying as a function of an angle~\cite{Barash78,Somers18}.
Changing the geometry and the composition of the system modifies the structure of the modes and their corresponding energy contribution, strongly affecting the sign and strength of the interaction~\cite{Boyer68,Mamaev79,Mamaev79b,Bordag01,Levin10,Graham13}. This means etermining and analyzing the electromagnetic modes allows to better understand the Casimir effect and can open pathways to tune it.

\subsection{A change of perspective}
\label{subsec:2}

In 1955 Lifshitz~\cite{Lifshitz56} formulated an alternative theory of the interaction between two plates. His theory was inspired by the work of Rytov on electrical fluctuations and thermal radiation~\cite{Rytov53} and by the experiment of Derjaguin and Abrikosova on the London-van der Waals forces between extended bodies~\cite{Derjaguin60}. Lifshitz' approach focused more on the force instead of the energy. Rather than evaluate the zero point energy, he considered the net radiation pressure exerted on one of the planes by the quantum and thermal electromagnetic fluctuations. The cornerstone of this procedure is the quantum average of the Maxwell stress tensor~\cite{Jackson75} operator, which is performed using the fluctuation-dissipation theorem~\cite{Kubo66,Marconi08}. 
This approach is conceptually different from that used by Casimir and much more related to the work of London~\cite{London30} on the interaction between two molecules. 
Although the zero point energy and quantum fluctuations are related concepts, the accent is put on the material dynamics and their quantum fluctuations, while the electromagnetic field, which was the main actor in Casimir's evaluation, is relegated to the role of the carrier of the interaction across the system. Neither the cavity modes nor their zero-point energy explicitly appear in the derivation leading to the famous Lifshitz formula~\cite{Lifshitz56}. For two parallel planes placed in vacuum and made of different materials the interaction only depends on the reflection coefficients $r^{\sigma}_{1}$ and $r^{\sigma}_{2}$ at the vacuum-material interface. At finite temperature, for planes separated by a distance $L$, the Lifshitz free energy is given by
\begin{align}
\mathcal{F}_{\rm Lif}=\mathrm{Im}\left\{\int_{0}^{\infty}\frac{d\omega}{2\pi}A\int\frac{d\mathbf{k}}{(2\pi)^{2}}\sum_{\sigma}\hbar\coth\left[\frac{\hbar\omega}{2k_{\rm B}T}\right]\;\ln\left[1-r^{\sigma}_{1}(\omega,k)r^{\sigma}_{2}(\omega,k)e^{-2\kappa L}\right]\right\}~,
\label{Elifshitz}
\end{align}
where $k=|\mathbf{k}|$ and $\kappa=\sqrt{k^{2}-\omega^{2}/c^{2}}$ ($\mathrm{Im}[\kappa]\le 0; \mathrm{Re}[\kappa] \ge 0$). Although Eq.~\eqref{Elifshitz} is valid for two generic parallel planar plates, in 1955 Lifshitz specifically considered two semi-infinite nonmagnetic bulks facing each other. In this case the reflection coefficients are given by the Fresnel formulae~\cite{Jackson75}
\begin{equation}
r_{i}^{\rm TE}(\omega, k)=\frac{\kappa-\kappa_{i}}{\kappa+\kappa_{i}},
\quad
r_{i}^{\rm TM}(\omega, k)=\frac{\epsilon_{i}(\omega)\kappa-\kappa_{i}}{\epsilon_{i}(\omega)\kappa+\kappa_{i}}~,
\label{eq:fresnel}
\end{equation}
where $\kappa_i=\sqrt{k^2-\epsilon_{i}(\omega)\omega^2/c^2}$ and $\epsilon_{i}(\omega)$ is the dielectric function describing the material comprising the plate ``$i$''.

Lifshitz' and Casimir's results appear different enough to inspire some doubts about their connection to the same physical phenomenon~\cite{Mehra67}.
A first confirmation of their equivalence is provided by the fact that Eq.~\eqref{Elifshitz} for $T=0$ and perfect conductors (i.e. $r_{i}^{\rm TE}=-1$ and $r_{i}^{\rm TM}=1$) reproduces Casimir's result in  Eq.~\eqref{EnergyForce}.  Differences appear as soon as more realistic materials are considered. For metals, the Lifshitz formula recovers Eq.~\eqref{EnergyForce} for sufficiently large values of $L$, while at short separations it predicts an energy scaling as $\propto L^{-2}$ instead of $L^{-3}$~\cite{Bordag00,Bezerra00,Genet00,Genet04}. This change in the exponent of the power law is similar to that discussed by Casimir and Polder on the influence of retardation on the London-van der Waals forces~\cite{Casimir48a,Intravaia11}. Indeed, it implicitly shows that Lifshitz' procedure includes a nonretarded (or quasi-static) interaction between the plates, which does not appear in Casimir expression in Eq.~\eqref{EnergyForce}.  

A clear physical understanding of this behavior is obtained when Casimir's and Lifshitz' approaches are connected and the change in the electromagnetic modes' structure, induced by real conductors, is analyzed. The link was pointed out by van Kampen et al. in 1968~\cite{Van-Kampen68} and further investigated by many other authors later~\cite{Ninham70,Renne71,Renne71a,Mahanty72,Langbein73,Langbein73a,Schram73,Ford85,Ford88,Bordag01,Bordag06,Intravaia12b}. 
The main idea relies on a corollary of the residue theorem called argument principle~\cite{Markusevic88}. If $f(z)$ is a monodromic function on a domain $D$, $\phi(z)$ an
analytic function on the same domain and $C$ a closed path contained in the domain $D$,
 the argument principle states that\cite{Markusevic88}
\begin{gather}
 \frac{1}{2\pi\imath}\oint_{C}
\phi(z)\partial_{z}\ln f(z) dz=\sum^m_{j=1}a_j\;\phi(z^{\rm zero}_j)-\sum^n_{j=1}b_j\;\phi(z^{\rm pole}_j)~,
\label{arguemPr}
\end{gather}
where $z^{\rm zero}_j$ and $z^{\rm pole}_j$ are respectively the zeros with multiciplity $a_j$ and the poles with multiplicity  $b_j$ of $f(z)$ contained in $C$. 
The connections between Lifshitz' and Casimir's approach can be made by recognizing that the frequencies of the electromagnetic field vibrating within a cavity with length $L$ are in general given by the solutions of 
\begin{equation}
1-r^{\sigma}_{1}(\omega,k)\; r^{\sigma}_{2}(\omega,k)\;e^{-2\kappa L}=0~.
\label{realModes}
\end{equation}
However, special attention must be devoted to the treatment of the branch cuts appearing in the previous expression~\cite{Schram73,Langbein73,Langbein73a,Intravaia12b}.
For two semi-infinite bulks comprised by dispersive and non dissipative materials the solutions of Eq.~\eqref{realModes} are all real~\cite{Schram73}. Using a contour which encircles the right half of the complex plane one can show~\cite{Intravaia12b} that at zero temperature\footnote{At finite temperature a generalization is also possible~\cite{Ninham70,Haakh10}.} Eq.~\eqref{Elifshitz} can be equivalently written in the form of Eq.~\eqref{sum-over-modes}.
Physically, this means that the fluctuation-induced force derived by Lifshitz is not only directly linked to the zero-point energy of each of the single mode frequencies vibrating within the cavity but importantly it also inherently includes the difference between configurations at $L$ and $L\to \infty$. 
Notice that Casimir's approach demands that all the frequency modes $\omega_{\mathbf{K}}$ are real in order to have a clear physical meaning. The equivalence with Eq.~\eqref{sum-over-modes} then necessarily imposes some restrictions on the behavior of the reflection coefficients, which are not needed by the Lifshitz formula. Indeed, as van Kampen~\cite{Van-Kampen68} and other authors~\cite{Schram73}, we have explicitly requested a dispersive but non dissipative form for the permittivity $\epsilon_{i}(\omega)$. 
Interestingly, although the previous arguments validate and connect both Lifshitz' and Casimir's approaches, they also point out some evident difficulties affecting the mode point of view but not the fluctuation perspective. This is particularly clear when dissipation in the system can no longer be neglected. As we will discuss in Secs.~\ref{sec:1-1} and \ref{OQS}, dissipation can appear in electromagnetic systems through different mechanisms and the frequency modes can become complex-valued functions even if the permittivity model is nondissipative.

Before analyzing this point more carefully, it is instructive to first investigate the novelties appearing in the modal analysis of the Casimir effect when one considers dispersive nondissipative materials instead of perfect electric conductors.

\subsection{Surface modes and the Casimir effect}
\label{nondissipativeModes}

\begin{figure}
\includegraphics[width=6cm]{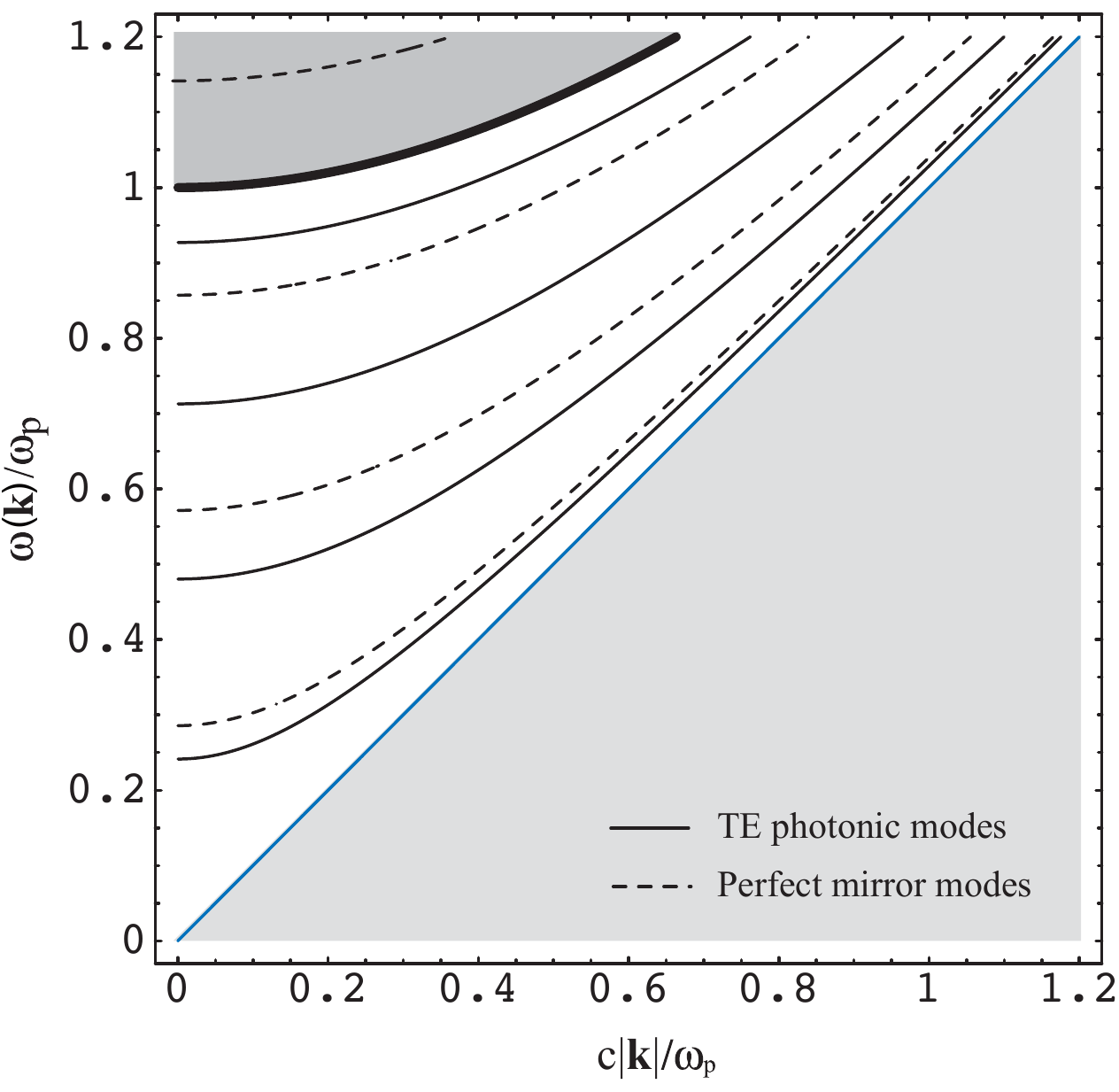}
\hspace{0.2cm}
\includegraphics[width=6cm]{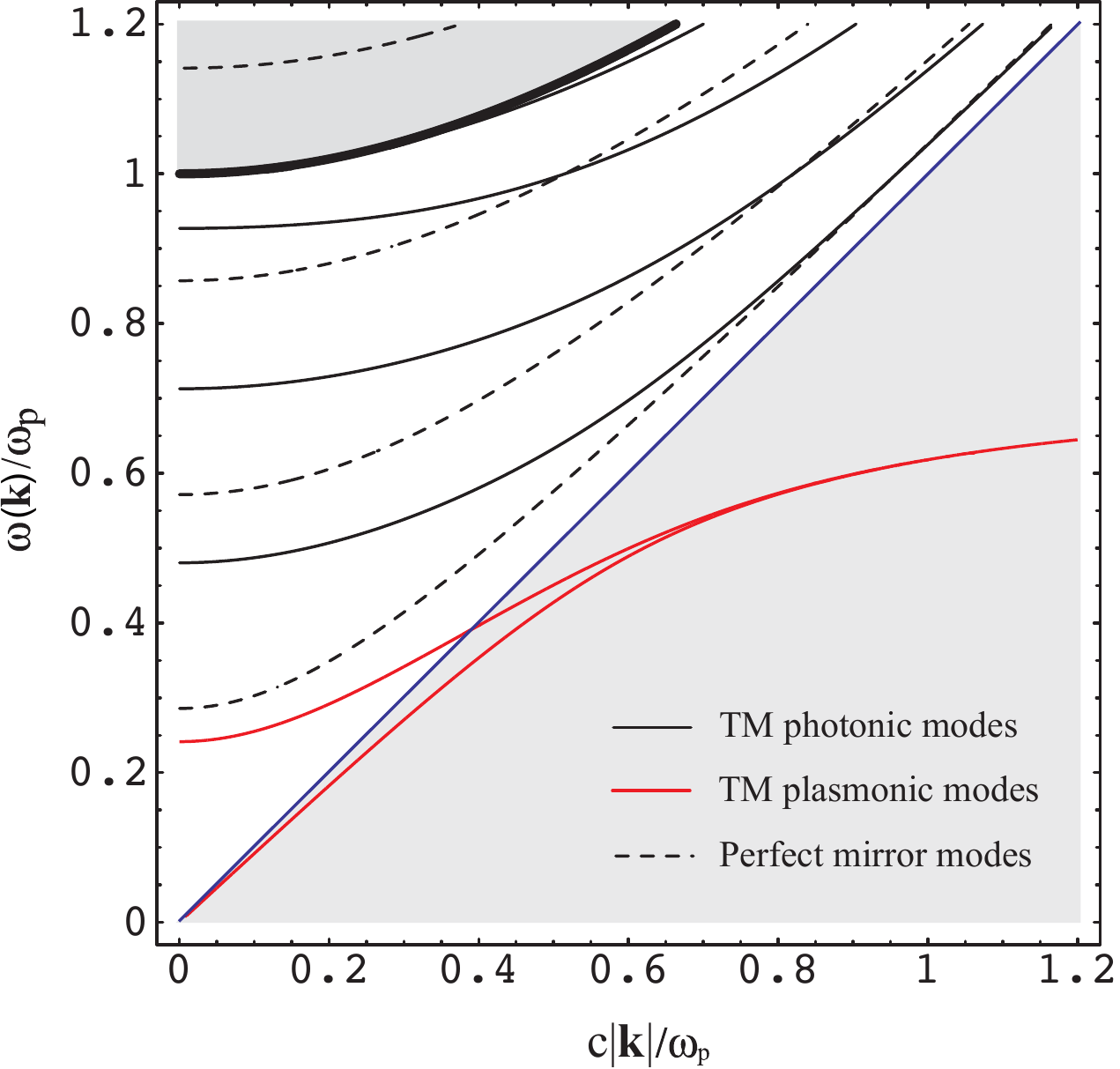}
\caption{Dispersion relations for TE- (left) and TM-polarized modes (right)
vibrating between two parallel semi-infinite bulks described by the plasma model (black solid line). The corresponding perfect conductor modes (dashed lines) are represented for comparison. In both cases $L = 1.75\,\lambda_{\rm p}$, where $\lambda_{\rm p}=2\pi c/\omega_{\rm p}$.
The diagonal line is the light cone
separating the evanescent (gray shadow) from the propagating region. Adapted from Ref.~\citen{Intravaia07}. 
}
\label{disptetm}
\end{figure}

Figure \ref{disptetm} shows the dispersion relations of the modes vibrating within a cavity formed by two identical nonmagnetic semi-infinite bulks described by the permittivity
\begin{equation}
\epsilon_{\rm Pl}(\omega)=1-\frac{\omega_{\rm p}^{2}}{\omega^{2}}.
\label{plasmaModel}
\end{equation}
The full lines are the solutions of Eq.~\eqref{realModes} (real valued functions), while the dashed lines describe the modes for perfect electric conductors. 
The permittivity in Eq.~\eqref{plasmaModel}, often referred to as the plasma model, is probably one of the simplest dispersive material models for a dielectric function~\cite{Jackson75}. It depends on just one parameter, the plasma frequency $\omega_{\rm p}$ and, although it is often used to describe a metal, it more precisely corresponds to a very simple description of a superconductor~\cite{Bimonte10} considered by F. London and H. London in 1935~\cite{London35} before the formulation of the BCS theory. 

\begin{figure}
\flushright
\parbox{5.7cm}{\includegraphics[width=5.4cm]{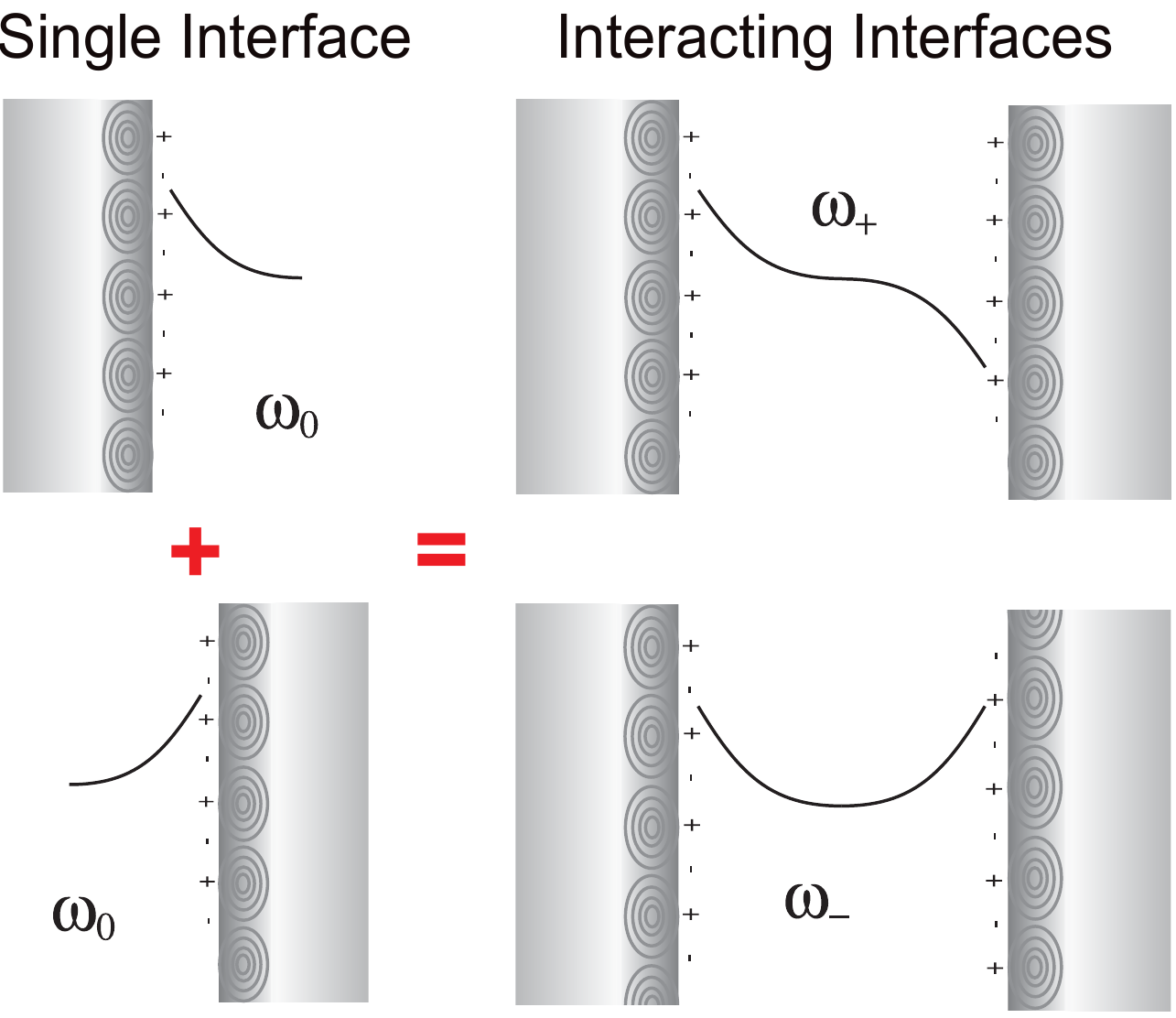}}
  \parbox{6.8cm}{\caption{
  When two vacuum/metal interfaces are sufficiently near to each other, the two respective surface plasmon-polaritons couple through their evanescent electromagnetic field. A frequency splitting occurs, giving rise to two new plasmonic modes. 
The Casimir force associated with the antisymmetric mode $\omega_{+}$ is an anti-binding force (repulsive), while the one for the symmetric mode $\omega_{-}$ is attractive.  Adapted from reference \cite{Intravaia07}. \label{pls}}}
\end{figure}

The introduction of dispersion in the dielectric response modifies the mode frequencies in various ways. We can distinguish three kinds of modes (see Fig.~\ref{disptetm}): Cavity modes -- propagating modes qualitatively similar to those obtained for perfect conductors; Bulk modes -- propagating modes mainly localized within the bulks; Evanescent modes -- having dispersion relations entering the region $\omega < c |\mathbf{k}|$ (light gray area below the light cone). Due to their propagating nature, we refer to both cavity and bulk modes  as ``photonic'' modes \cite{Intravaia07,Intravaia07a}. In the specific case of the permittivity in Eq.~\eqref{plasmaModel}, we have only two  TM-polarized ($\omega_{\pm}$) evanescent modes and we refer to them as ``plasmonic'' modes, since they result from the hybridization of the two surface plasmon-polariton modes existing at the vacuum-material interface of the two planes~\cite{Joulain05,Intravaia05,Intravaia07} (see Fig.~\ref{pls}).
Surface plasmon-polaritons~\cite{Joulain05,Maier07} are mixed light-matter modes involving collective electronic excitations coupled with an evanescent electromagnetic field. They belong to a class of solutions of Maxwell equations often also called surface polaritons, which appear due to the confinement of light-matter interaction at the boundary of an object. Other examples are surface phonon-polaritons~\cite{Joulain05} and surface magnon-polaritons~\cite{Matsuura83,Sloan19}.
Due to the connection with the electronic charge, the field of surface plasmon-polaritons is mainly electric in nature and characterizes the near field behavior at the interface to a conductor. This feature is connected to the TM-polarization, which has an electric field component perpendicular to the surface.

In the large distance limit, the two plasmonic modes must tend to the expression for the single interface surface plasmon-polariton mode. Within a local description, for a planar surface comprised by a nonmagnetic material, the dispersion relation of the surface plasmon-polariton is the solution of~\cite{Joulain05}
\begin{equation}
k=\frac{\omega}{c}\sqrt{\frac{\epsilon(\omega)}{\epsilon(\omega)+1}}
\label{surfaceMode}
\end{equation}
and for the plasma model this gives
\begin{equation}
\lim_{L\to \infty}\omega_{\pm}=\omega_{0}\equiv\omega_{\rm sp}=\omega_{\rm p}\,\sqrt{\left(\!\frac{ck}{\omega_{\rm p}}\!\right)^{\!\! 2}+\frac{1}{2} - \sqrt{\left(\!\frac{ck}{\omega_{\rm p}}\!\right)^{\!\! 4}+\frac{1}{4}}} \xrightarrow{k\to \infty}\frac{\omega_{\rm p}}{\sqrt{2}}~.
\label{sp_dispersion}
\end{equation}

At zero temperature the free energy is identical to the energy ($\mathcal{F}_{\rm Lif}\xrightarrow{T\to0}E_{\rm Lif}$) and, using the argument principle, we can rewrite the Lifshitz formula in Eq.~\eqref{Elifshitz} 
as the sum of two different contributions~\cite{Intravaia05,Intravaia07,Intravaia07a,Haakh10}
\begin{align}
\label{start} 
E_{\rm Lif}&=\underbrace{ 
\sum_{\mathbf{k}}\left[\frac{\hbar\omega_+}{2}
+\frac{\hbar\omega_-}{2}\right]^L_{L\rightarrow\infty}}_{\text{plasmonic
modes ($E_{\mathrm{pl}}$)}}
+\underbrace{
\sum_{\mathbf{K}}\left[
\frac{\hbar\omega^{\sigma}_n}{2}\right]^L_{L\rightarrow\infty}
}_{\text{photonic modes
($E_{\mathrm{ph}}$)}}~.
\end{align}

\begin{figure}
\parbox{7cm}{\includegraphics[width=6.8cm]{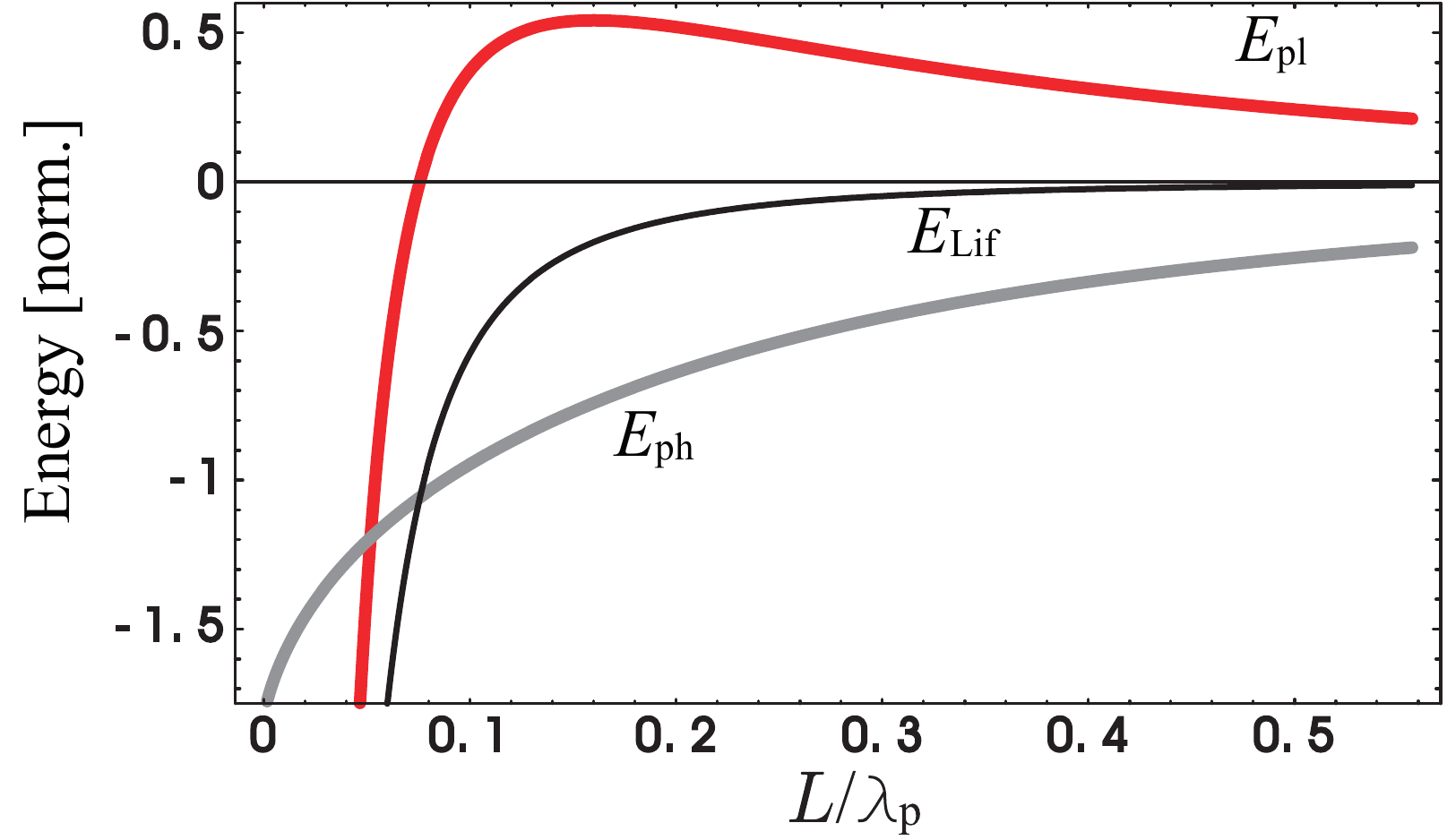}}
\parbox{5.5cm}{\caption{A plot of the normalized plasmonic, photonic, and total Casimir  energy ($E_{\rm pl}$, $E_{\rm ph}$ and $E_{\rm Lif}$ respectively) as a function of the separation between the plates.
We see that $E_{\rm pl}$ shows a maximum for $L/\lambda_{\rm p}\sim 0.16$ ($F_{\rm pl}$ changes its sign) while $E_{\rm ph}$ monotonically tends to zero. $F_{\rm ph}$ is always attractive
\cite{Intravaia05}. Adapted from Ref.~\citen{Intravaia07}. \label{nrgcotr}}}
\end{figure}

\begin{figure}
\centering
 \parbox{7cm}{\includegraphics[width=6.8cm]{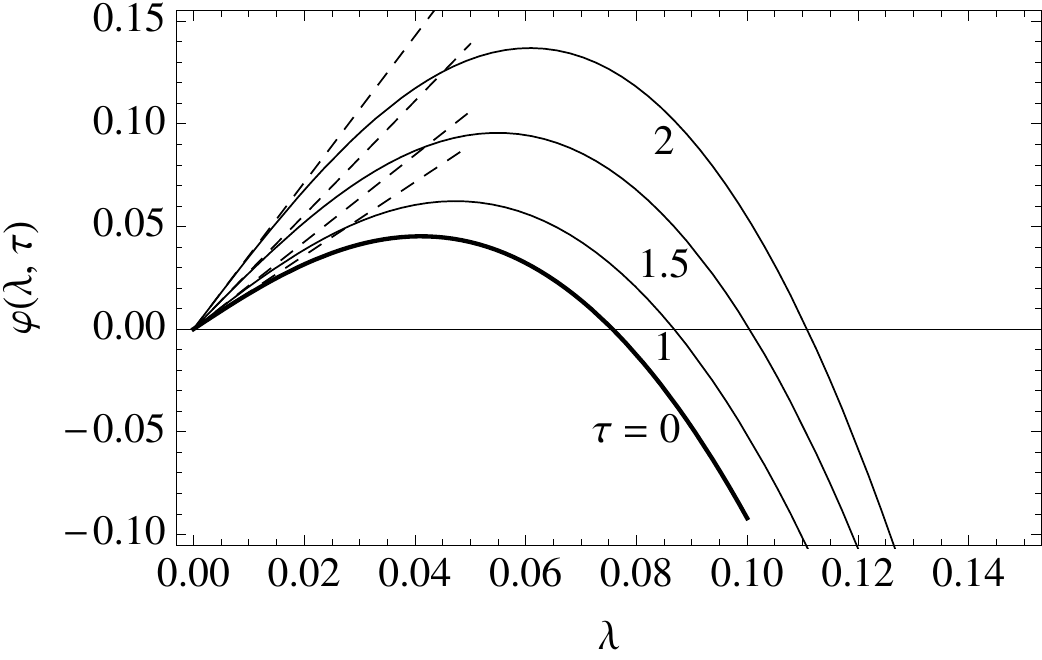}}
   \parbox{5.5cm}{ \caption{Plasmonic contribution to the Casimir free energy
     vs.\ distance at different temperatures, 
     normalized to the perfect mirror case at $T = 0$ ($\varphi=\mathcal{F}_{\rm pl}/E_{\rm Cas}$).
Distance and temperature are scaled to the plasma wavelength and temperature $T_{\rm p} 
= \hbar \omega_{\rm p} / k_B$, i.e. $\lambda=L/\lambda_{\rm p}$ and  $\tau=k_{\rm B}T/(\hbar\omega_{\rm p})$. 
Adapted from Ref.~\citen{Haakh10}.
\label{phi}}}
\end{figure}

Although the only observable is the total Casimir energy, 
evaluating the different contributions separately reveals striking features characterizing the underlying physics of the system.
For separations $L$ smaller than the plasma wavelength $\lambda_{p}=2\pi c/\omega_{\rm p}$, the Casimir effect has an interpretation which bridges the gap between quantum field theory and condensed matter physics. As pointed out by van Kampen~\cite{Van-Kampen68}, in this regime the Casimir force can be in fact understood as mainly resulting from the quasi-electrostatic interaction between the surface plasmon-polaritons. In our notation this is equivalent to write that for $ L\ll \lambda_{\rm p}$  one has $E_{\rm Lif}\sim E_{\mathrm{pl}}$, while $E_{\rm ph}$ can be neglected. Physically, the sub-leading nature of $E_{\rm ph}$ can be related to the high frequency transparency of the material comprising the bulks. Indeed, the frequency of the photonic modes scales as the inverse of $L$ and, since $\epsilon(\omega)\xrightarrow{\omega\gg \omega_{\rm p}} 1$, for sufficiently small separations these modes can no longer be sustained within the cavity formed by the two semi-infinite bulks.
Figure~\ref{nrgcotr} shows the two contributions in Eq.~\eqref{start} and the total Casimir energy. In particular, one has that~\cite{Genet04,Intravaia05,Intravaia07}
\begin{equation}
E_{\mathrm{pl}}\sim \alpha \frac{L}{\lambda_{\rm p}} E_{\rm Cas}\quad (\alpha=1.790)
\end{equation}
directly connecting the nonretarded limit of the Lifshitz formula with the interaction between the surface plasmon-polaritons. Since these surface modes exist only for dispersive materials, this also explains why the $\propto L^{-2}$ behavior does not appear in Casimir's calculation with perfect reflectors.
Figure~\ref{nrgcotr} also shows that for $L \gtrsim\lambda _{\mathrm{p}}$, $E_{\rm pl}$ has a much greater importance than one could have expected. At large distances, despite their evanescent nature, the two plasmonic modes give rise to a positive energy contribution having a magnitude much larger than the total Casimir force~\cite{Intravaia05,Intravaia07,Intravaia07a,Haakh10}. Interestingly, the slope of $E_{\rm pl}$ changes sign as a function of the distance $L$, and the corresponding force $F_{\rm pl}$, stemming from the plasmonic contribution, goes from attractive at short separations to repulsive at large ones. This means that at large separations the total interaction results from a fine balance between the repulsive plasmonic contribution and the slightly larger attractive interaction arising from the remaining photonic modes.  As shown in Fig.~\ref{phi} this outcome is also robust against a generalization of Eq.~\eqref{start} to finite temperatures~\cite{Haakh10}.

Although $F_{\mathrm{pl}}$ cannot be observed by itself, its repulsive nature at sufficiently large values of $L$ has led to suggestions aiming to tune the strength of the total Casimir force. 
These proposals are based on a tailoring of the polaritonic modes, using for example geometric nanostructuring \cite{Intravaia12a,Intravaia13,Davids14,Rodriguez13} or non-equilibrium configurations \cite{Haakh10}.
In particular, in nanophotonics the relevance of nanostructured surfaces with tailored plasmonic dispersion relations has already been demonstrated in many setups and applications ranging from extraordinary light transmission to surface-enhanced Raman scattering \cite{Ebbesen98,Nie97}. 
Concerning the Casimir interaction, an experiment performed in 2012 measured a nontrivial modification of the force between a nanostructured gold grating with dimensions smaller than plasma wavelength for gold ($\lambda_{\rm p}^{\rm Au}\sim140$ nm) and a gold-coated sphere of radius $150 \mu$m~\cite{Intravaia13}.
A new behavior in the Casimir interaction was observed which is significantly different from the well-known attraction with an unstructured plate. It is characterized by a crossover from an enhancement to a strong reduction of the Casimir force which depends on the period of the grating. In addition, at large inter-surface separations, the Casimir interaction decreases faster than the usual $L^{-4}$ power law. This behavior at least qualitatively correlates with what would be expected by an enhancement of the plasmonic contribution. However, state-of-the-art theoretical modeling, based on the proximity force approximation~\cite{Fosco12,Bimonte17,Bimonte21} for treating the curvature of the large-radius sphere and an exact ab-initio scattering analysis of the resulting effective plane-grating geometry, was not able to reproduce the experimental findings~\cite{Davids10,Intravaia12a,Intravaia13}. The development of an analytical or numerical analysis of the sphere-grating geometry, capable of dealing with the disparate length scales present in the setup, remains an open problem.

Finally, it is interesting to mention that the relevance of surface polaritons for the Casimir interaction between two surfaces is not limited to conductors. A qualitatively similar but physically richer behavior than in Fig.~\ref{nrgcotr} has been observed for a magnetodielectric cavity~\cite{Haakh13}. Even the zero temperature interaction between two parallel graphene layers can be better understood by investigating the modes existing in the system \cite{Egerland19}.  Differently from usual conductors, in graphene, surface polaritons can appear also in the TE-polarization~\cite{Mikhailov07,Bordag14,Bordag15,Werra16a,Werra16,Egerland19}.

\section{Dissipation, fluctuations and zero-point energy}
\label{sec:1-1}

Casimir's sum over modes approach provides a useful tool in order to reveal and investigate the underlying physics of the different modes' contributions participating in the interaction. 
However, despite its interest, when compared to Lifshitz' approach, it has some serious limitations. In realistic situations, due to the presence of some form of dissipation, modes turn into resonances (quasi-normal modes~\cite{Kristensen20}), which are described by complex valued frequencies. Mathematically, when they are inserted in Eq.~\eqref{sum-over-modes}, one cannot \emph{a priori} guarantee that the resulting expression is real valued, making its interpretation as an energy difficult. Physically, the exchange of energy between the system and its surroundings typical of dissipative interactions prompts the need for the generalization of the idea of a mode and consequently of our understanding of the zero-point energy.

In quantum mechanics dissipation is usually introduced within the so-called \emph{open quantum system} paradigm~\cite{Breuer02,Hanggi05}. It amounts to enlarging the Hilbert space of the system of interest, coupling it with (at least) one additional larger system, generically called \emph{bath} or \emph{environment}. The specific physical nature of the environment might be to some extent quite diverse (coupled oscillators or spins~\cite{Feynman63,Ford65,Caldeira81,Breuer02} or also a gas of particles~\cite{Ciccarello17}) but it is usually comprised of a large number of degrees of freedom. 
The logic behind this extension is twofold. First, it allows for an exchange of energy between the system and the environment. 
Secondly, following an argument typical of statistical mechanics, a large number of the environment's degrees of freedom allows for a very large Poincaré recurrence time and for a net flux of energy from the system towards the environment that we can interpret as dissipation. 

In order to further clarify the previous concepts it is convenient to analyze a basic and completely solvable example, i.e. a single harmonic oscillator with frequency $\omega_{\rm a}$ (the system of interest) coupled to a bath of harmonic oscillators with frequencies $\omega_{j}$ (its environment)~\cite{Intravaia03,Intravaia03a,Maniscalco04b,Hanggi05}. A typical Hamiltonian for this system is given by \cite{Ford87a,Ford88a}
\begin{equation}
\label{coupledH}
\hat{H}=\frac{1}{2}\left(\hat{p}_{a}^{2}+\omega^{2}_{\rm a}\hat{q}_{a}^{2}\right)+\sum_{j=1}^{\infty}\frac{1}{2}\left(\hat{p}_{j}^{2}+\omega^{2}_{j}[\hat{q}_{j}-\hat{q}_{0}]^{2}\right)~,
\end{equation}
where we assumed all the masses to be equal to unity. Due to the interaction with the bath, the minimal energy of the single oscillator is no longer $E_{0}=\hbar \omega_{\rm a}/2$ but it can be defined as the difference between the energy of the coupled system and the energy of the bath alone~\cite{Ford85,Ford88,Hanggi06}. At zero temperature one has
\begin{equation}
\tilde{E}_{0}=\sum_{i=0}^{\infty}\frac{\hbar\varpi_{i}}{2}-\sum_{j=1}^{\infty}\frac{\hbar\omega_{j}}{2}~,
\end{equation}
where $\varpi_{i}$ are the eigenfrequencies obtained by diagonalizing the system of oscillators in Eq.~\eqref{coupledH}. Instead of actually diagonalizing the Hamiltonian it is more convenient to consider the susceptibility of the single oscillator obtained by a self-consistent treatment of its equation of motion~\cite{Ford85,Ford88,Ford88a}. In the frequency domain one can show that the motion of the single oscillator is described by
\begin{equation}
\hat{q}_{a}(\omega)=\alpha(\omega)\hat{f}(\omega)
\end{equation}
where $\hat{f}(\omega)$ is connected to the free evolution of the bath's oscillators \cite{Ford88a,Intravaia12b}, while
\begin{equation}
\alpha(\omega)= \frac{1}{\omega^{2}_{\rm a} -\omega^{2}-\imath \omega \mu(\omega)},\quad \mu(\omega)= - \imath \omega\sum_{j = 1}^{\infty}\frac{\omega^{2}_{j}}{\omega^{2}_{j}-\omega^{2}}
\label{DissAlpha}
\end{equation}
is the test-oscillator's susceptibility. From the previous expression, one can deduce that the frequencies of the bath are the zeros of the polarizability. Also by definition, the modes of the coupled system are those frequencies for which a motion can be sustained without any external drive. This means that $\hat{q}_{a}(\omega)$ can be nonzero also if $\hat{f}(\omega)=0$. This indicates that the $\varpi_{i}$ are the poles of $\alpha(\omega)$~\cite{Ford85,Ford88}. Moving onto the complex plane $\omega\to \zeta$, choosing a contour $C$ encircling the positive frequency axis and using Eq.~\eqref{arguemPr} we can write~\cite{Renne71,Ford85}
\begin{equation}
\tilde{E}_{0}=- \frac{1}{2\pi\imath}\oint_{C}d\zeta
\frac{\hbar \zeta}{2}\partial_{\zeta}\ln [ \zeta^{2}\alpha(\zeta)] = \int_{0}^{\infty}\frac{d\omega}{2\pi}\;\hbar \omega\;\mathrm{Im}\left\{\partial_{\omega}\ln [\omega^{2}\alpha(\omega+\imath 0^{+})]\right\}~. 
\label{etilde}
\end{equation}
Writing the last equation is equivalent to choosing a causal expression for the susceptibility, which is consequently assumed to be analytic in the upper
part of the complex plane~\cite{Renne71,Ford85}.
If we take the limit for which the bath's mode-frequencies form a continuum, we obtain
\begin{equation}
\label{Cont-Limit}
\mu(\omega+\imath 0^{+})
\to - \frac{\imath}{2} (\omega+\imath 0^{+})\int_{-\infty}^{\infty}{\rm d}\nu \;\rho(\nu)\frac{\nu^{2}}{\nu^{2}-(\omega + i0^+)^{2}}~,
\end{equation}
where we have introduced the bath density of modes $\rho(\nu)$.
In the last equation we also extended the $\nu$-integral from $-\infty$ to $+\infty$ requiring that the product of $\rho(\nu)\nu^{2}$ is even in $\nu$.
The expression in Eq.~\eqref{etilde} is more general than the specific model defined by the Hamiltonian in Eq.~\eqref{coupledH} and occurs for other descriptions of the environment and couplings to the system of interest. General physical principles impose that $\mu(\omega)$ is a ``positive function'' \cite{Ford88a}, which means that (i) it is analytical in the upper-half of the complex-frequency plane (causality condition), (ii) it has a non negative real part at the upper boundary of the real axis (to preserve the second law of thermodynamics), and (iii), given the complex frequency $\zeta$, it must satisfy the crossing relation $\mu(\zeta)=\mu^{*}(-\zeta^{*})$ (reality of the function in the time domain) \cite{Ford88a,Tatarskii87}.
It is interesting to remark that Eq.~\eqref{etilde} can be obtained proceeding as Lifshitz, focusing on the test system's fluctuations and using the fluctuation-dissipation theorem~\cite{Ford88}. 

A typical expression for $\rho(\nu)\nu^{2}$, which fulfills the requirements imposed by the physics of the system is~\cite{Weiss08}
\begin{equation}
\rho(\nu)\nu^{2}=\frac{2\Gamma}{\pi}
\frac{1}{1+\nu^{2}\tau_{c}^{2}},\quad (\Gamma\text{ positive})
\end{equation}
where $\tau_{c}$ is connected with the so-called bath's response time~\cite{Weiss08}. This leads to the following form for the susceptibility
\begin{equation}
\alpha(\omega)=\frac{1}{\omega^{2}_{\rm a}-\omega^{2}-i \omega\frac{\Gamma}{1-\imath \omega\tau_{c}}}~,
\label{DrudePolar}
\end{equation}
which for $\tau_{c}=0$ corresponds to the expression one would expect from a damped harmonic oscillator with damping rate $\Gamma$. 
Notice that the bath's mode-structure seems to disappear in the previous expression and that for $\tau_{c}=0$ the susceptibility features only two \emph{resonance} poles in $\Omega_{1}=\Omega_{\rm a}$ and $\Omega_{-1}=-\Omega_{\rm a}^{*}$, where we defined
\begin{equation}
\Omega_{\rm a}=\sqrt{\omega_{\rm a}^{2}-\frac{\Gamma^{2}}{4}}-\imath \frac{\Gamma}{2}~,
\label{complexClassical}
\end{equation}
i.e. the usual resonances of a damped harmonic oscillator~\cite{Weiss08}.
However, the choice $\tau_{c}=0$ corresponds to a rather idealized and to some extent unphysical situation. It also leads to divergencies in the calculation of diverse quantities~~\cite{Weiss08}. Physically, for $\tau_{c}>0$ the properties of a damped harmonic oscillator are still recovered as long as $\omega_{a}\tau_{c}\ll1$. However, for $\tau_{c}>0$, $0<\omega_{a}\tau_{c}\ll 1$ and $\omega_{a}>\Gamma/2$, Eq.~\eqref{DrudePolar} features a zero at $\zeta=-\imath/\tau_{c}$ and three poles: Two are complex conjugated with nonzero real parts and negative imaginary parts. They are generalizations of those given before Eq.~\eqref{complexClassical} and we will still call them $\Omega_{1}$ and $\Omega_{-1}$. The third pole at $\Omega_{0}\equiv-\imath \xi_{0}$ is purely imaginary ($\xi_{0}$ positive) and located on the negative imaginary frequency axis \cite{Hanke95,Weiss08}, corresponding to an overdamped oscillation. 

With this information, using Eq.~\eqref{DrudePolar} we can analytically evaluate Eq.~\eqref{etilde}
\begin{equation}
\tilde{E}_{0}=\sum_{m=-1}^{1}\frac{\hbar}{4}\left( \Omega_{m}-\frac{2\imath}{\pi}\Omega_{m}\ln[\Omega_{m}\tau_{c}]\right)
=\sideset{}{'}\sum_{m=0}^{1}\frac{\hbar}{2}\mathrm{Re}\left[ \Omega_{m}-\frac{2\imath}{\pi}\Omega_{m}\ln[\Omega_{m}\tau_{c}]\right]
\label{dissEnergy}
\end{equation}
where the prime in the sum indicates that the addend with $m=0$, corresponding to the purely imaginary pole, must be weighted with $1/2$. Since the denominator of Eq.~\eqref{DrudePolar} is proportional to a polynomial of the third order, due to Vieta's relations one has that
\begin{equation}
\sum_{m=-1}^{1}\Omega_{m}=-\frac{\imath}{\tau_{c}} \quad \text{or also}\quad \sideset{}{'}\sum_{m=0}^{1} \mathrm{Im}\left[\Omega_{m}\right]=-\frac{1}{2\tau_{c}}~,
\label{sumRule}
\end{equation}
showing that the imaginary part of the poles sum up to a constant, the zero of the polarizability, that is exclusively related with the properties of the bath as an isolated entity, i.e. decoupled from the oscillator.

At this point a few comments are in order. 
The expression in Eq.~\eqref{dissEnergy} highlights that for a quantum dissipative oscillator the minimal energy is rather nontrivial. While even classically one would have expected a complex resonance as in Eq.~\eqref{complexClassical} for the damped oscillator, a naive generalization of its quantum minimal energy $\tilde{E}_{0}$ as $\hbar \mathrm{Re}[\Omega_{a}]/2$ is for different reasons incorrect~\cite{Langbein70,Langbein73,Langbein73a}. Indeed, in addition to the appearance of additional resonances, $\tilde{E}_{0}$ features a logarithm which can be associated with the contribution of the bath fluctuations to the minimal energy~\cite{Nagaev02,Intravaia12b}.
Notice that before taking the continuum limit for the bath spectrum the susceptibility is even in frequency (micro-reversibility). However, this is no longer true for Eq.~\eqref{DrudePolar} which, as any generic susceptibility, must also fulfill the crossing relation $\alpha(\zeta)=\alpha^{*}(-\zeta^{*})$. 
It is the latter that in general prescribes the poles' and zeros' structure of the polarizability. Indeed, because of the crossing relation,  if $\Omega$ is a pole or a zero then $-\Omega^{*}$ must be one as well.

\begin{figure}[t]
\begin{center}
\parbox{6.7cm}{\includegraphics[width=6.5cm]{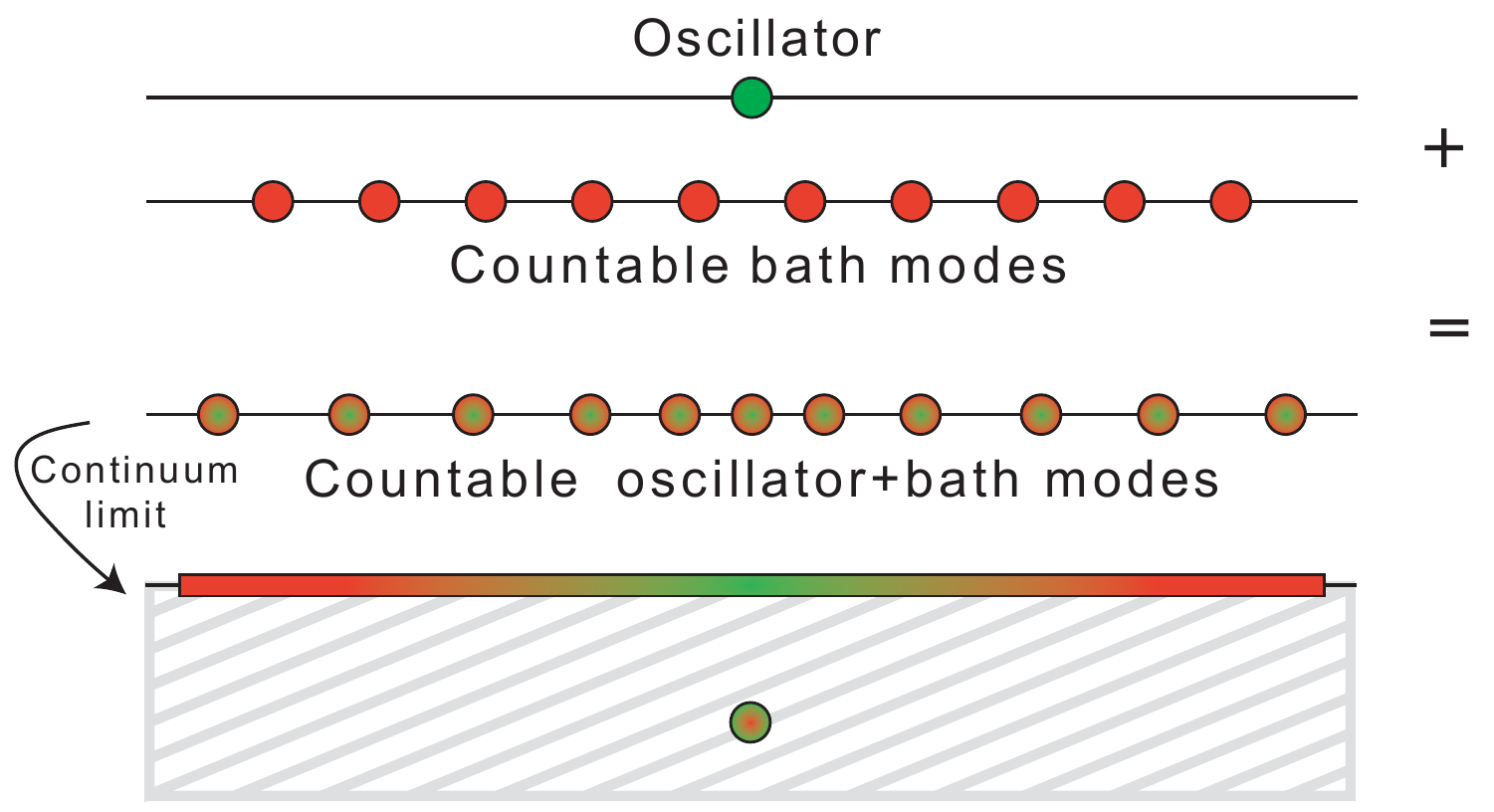}}
\parbox{5.8cm}{\caption{Schematic representation of a harmonic oscillator (system of interest) coupled to a collection of harmonic oscillators (bath). The modes of the isolated system (green dot) and of the isolated bath (red dots) are described by countable real frequencies. The same applies to the modes of the coupled system (green-red dots). In the continuum limit the mode spectrum becomes dense giving rise to a singularity along the whole real axis and leads to the appearance of a complex pole on the lower half of the complex plane. Adapted from Ref.~\cite{Intravaia12b}}
\label{polecont}}
\end{center}
\end{figure}

Strictly speaking, independently from the choice of $\rho(\nu)\nu^{2}$, the expression for the susceptibility entering in Eq.~\eqref{etilde} is per construction only valid in the upper part of the complex frequency plane~\cite{Renne71,Ford85,Sernelius06,Intravaia12b}.  Considering the singularities of Eq.~\eqref{DrudePolar}  with negative imaginary part is therefore inherently related to an analytic continuation of the polarizability to the lower half of the complex plane. This occurs across a discontinuity located along the real frequency axis, which is generated by the coalescence of the bath spectrum in the limit of continuous mode-frequencies (see Fig.~\ref{polecont}). The complex poles are then located on a different, to some extent  ``unphysical'', Riemann sheet~\cite{Ford85} with respect to the expression we would have obtained starting from Eq.~\eqref{etilde} but with $\alpha(\omega-\imath 0^{+})$. The sum-rule in Eq.~\eqref{sumRule} itself can be seen as a consequence of this analytical continuation. In general, since causality always requires that all the poles\footnote{Here we assume for simplicity that the polarizability is a meromorphic function with poles as the only discontinuities.} and the zeros of the polarizability in Eq.~\eqref{DissAlpha} must be located in the lower half of the complex plane one has that
\begin{equation}
0=-\frac{1}{2\pi \imath}\int_{-\infty}^{\infty}\frac{d\omega}{2\pi}\; \omega\; \partial_{\omega} \ln\left[\omega^{2}\alpha(\omega+\imath 0^{+})\right]=\sum_{i}\Omega_{i}^{\rm pole}-\sum_{j}\Omega_{j}^{\rm zero}~.
\label{sumRuleSimple}
\end{equation}
This is the generalization of Eq.~\eqref{sumRule}, where $\Omega_{i}^{\rm pole}$ solve $\omega^{2}_{\rm a} -\omega^{2}-\imath \omega \mu(\omega)=0$, while $\Omega_{j}^{\rm zero}$ are the solutions of $1/\mu(\omega)=0$ and therefore determined by the uncoupled bath.
The zero on the left of Eq.~\eqref{sumRuleSimple} is obtained using the argument principle [Eq.~\eqref{arguemPr}] and choosing a contour encircling the upper part of the complex plane. The expression on the right of Eq.~\eqref{sumRuleSimple} is instead obtained using the argument principle with a contour encircling the lower part of the complex plane and the analytically continued expression of the polarizability.

\section{Casimir energy within the open quantum system approach} 
\label{OQS}

The previous discussion highlights that dissipation affects the ground state energy of a quantum mechanical system in a nontrivial way. 
Similar modifications must be expected for each of the modes considered in the calculation of the Casimir energy. They are hidden, however, behind the elegance of Lifshitz' approach. Despite the prominent role played by the fluctuation-dissipation theorem, for the Lifshitz formula in Eq.~\eqref{Elifshitz} the effect of dissipation is not immediately evident because it is encoded in the expressions for the reflection coefficients. 
A `sum over modes' approach is nevertheless possible and one can indeed show~\cite{Intravaia08,Intravaia12b} that the Lifshitz formula is equivalent to
\begin{equation}
\label{eq:Casimir-Diss}
E_{\rm Lif} = \frac{\hbar}{2} 
 \mathrm{Re}\,\left[\sideset{}{'}\sum_{\mathbf{K}}
\Omega_{m}-
\frac{2\imath\Omega_{m}}{\pi}
\ln\left[\Omega_{m}\tau_c\right]\right]^{L}_{L\to\infty}
, \qquad
\mathrm{Im}\left[\sideset{}{'}\sum_{\mathbf{K}}
\Omega_{m}\right]_{\to \infty}^{L}=0
\end{equation}
where $\Omega_{m}$ are complex frequencies and
the prime indicates that terms for which $\Omega_{m}$ is an imaginary number are weighted 
with $1/2$. The expression in Eq.~\eqref{eq:Casimir-Diss} is the generalization of Eq.~\eqref{sum-over-modes}. Notice that, similarly to the single harmonic oscillator, the energy is neither the sum over the complex modes of the system nor over their real parts as suggested in some previous works~\cite{Langbein70,Zhou95,Sernelius06}.
As above, within the open quantum system approach~\cite{Weiss08}, the logarithmic correction in Eq.~\eqref{eq:Casimir-Diss} can be understood as arising from the fluctuations of the bath~\cite{Nagaev02}. With respect to the results of the previous section, the sum rule on the right of Eq.~\eqref{eq:Casimir-Diss} has, however, a wider impact on the system's energy.
Indeed, although for dimensional reasons and analogy to the dissipative harmonic oscillator a response time $\tau_c$ must appear in the expression, due to the sum rule, the total Casimir energy is independent from this constant. In particular, the sum rule in Eq.~\eqref{eq:Casimir-Diss} can be seen as resulting from the fact that the dissipative bath as an isolated entity does not change its properties when the distance changes from $L$ to $L\to \infty$.

Despite the fact that the open quantum system approach provides a quite general framework for describing dissipation, it is also important to understand the physical differences among the possible mechanisms of energy loss. If light-matter interaction is present, one can roughly distinguish two of them. The first, which we call intrinsic dissipation, is connected with the properties of the involved materials and their number of internal degrees of freedom as well as their interaction. A prominent example is Ohmic dissipation due, for example, to impurities, electron-electron or electron-photon interaction. The electromagnetic field itself is responsible for the second kind of dissipation. In addition to mediate the interaction between the objects within the system, it can directly play the role of the bath, carrying energy away in the form of radiation. For simplicity, we call this induced dissipation but one usually also speaks of radiative damping. A basic example is spontaneous decay in atomic systems. Induced dissipation depends on the boundary conditions that the electromagnetic field has to fulfill when it interacts with the objects in the system. As such, it depends on the geometry of the system and it can be enhanced or suppressed, as for example in periodic structures (e.g. photonic crystals\cite{Yablonovitch87,John87,John90,Vats02}).

Despite the complication induced by the introduction of dissipation, the sum over modes approach 
is still able to provide an `anatomic view' of the Casimir effect, allowing for the analysis of different contributions and their physical origin. 
To illustrate the utility of Eq.~\eqref{eq:Casimir-Diss}, we consider in the following two relevant examples.

\subsection{Dissipative Casimir interaction at short distances}

The first example is a generalization of the plasmonic interaction analyzed in Sec. \ref{nondissipativeModes}.
We stick with the same geometry considered by Lifshitz, i.e. two parallel semi-infinite identical bulks separated by vacuum but, instead of plasma model in Eq.~\eqref{plasmaModel}, we  use here
\begin{equation}
\epsilon_{\rm D}(\omega)=1-\frac{\omega_{\rm p}^{2}}{\omega(\omega+\imath \gamma)}
\label{DrudeModel}
\end{equation}
where $\gamma$ is the phenomenological damping rate used to quantify the intrinsic dissipation in the conductor.
The model in Eq.~\eqref{DrudeModel} often goes in the literature under the name of Drude model \cite{Jackson75}. 

For the geometry considered by Lifshitz, due to the boundary conditions, only intrinsic dissipation plays a role in the calculation. The energy cannot be radiated away from the system but only exchanged with the bath coupled with the internal degrees of freedom the conducting material. 
As in the non dissipative case, the dispersion relations are given by the solutions of  Eq.~\eqref{realModes}. One finds two plasmonic modes $\Omega_\pm( k )$ originating from the interaction between the surface plasmon-polaritons. In analogy to Eq.~\eqref{start} and in connection with Eq.~\eqref{eq:Casimir-Diss}, we define the plasmonic contribution to the Casimir energy as 
\begin{equation}
E_{\rm pl} = \frac{\hbar A }{2} 
\int\!\frac{ k {\rm d}k }{ 2\pi } 
\mathrm{Re}\,\sum_{i=\pm}
\left[\Omega_{i}(k) -
\frac{2\imath}{\pi}\Omega_{i}(k)
\ln[\Omega_{i}(k)\tau_c]
\right]^{L}_{L\to\infty}~.
\label{smalldiss1}
\end{equation}
At a distance smaller than the plasma wavelength  $\lambda_{\rm p}$ the dissipative dispersion relation for the coupled polaritonic modes are given by a slight generalization of the dispersion relations determined by van Kampen~\cite{Van-Kampen68} 
\begin{equation}
\Omega_{\pm} = 
	\sqrt{\omega^2_{\pm}-\frac{\gamma^2}{4}}
	- \imath\frac{\gamma}{2}
,\qquad 
\omega^2_{\pm} = \frac{ \omega^2_{\rm p} }{ 2 }
\left(1\pm e^{ -kL}\right).
\end{equation} 
One can check that in this limit the sum rule in Eq.~\eqref{eq:Casimir-Diss} is automatically satisfied. 
To leading order in $\gamma/\omega_{\rm p}$ (good conductors) Eq.~\eqref{smalldiss1} yields
\begin{equation}
	E_{\rm pl} \approx 
	-\frac{\pi^2 \hbar c A}{720 L^3}
\frac{3}{2}\left[\alpha \frac{L}{\lambda_{\rm p}}
-
\frac{15\zeta(3)}{\pi^4} 
\frac{\gamma L}{ c }
\right]
,\qquad
\alpha=1.193\ldots
	\label{eq:short-distance-expansion}
\end{equation}
where for the value of the The Riemann zeta function one has $\zeta(3) \approx 1.202$.
The expression in Eq.~\eqref{eq:short-distance-expansion} coincides with the short distance limit of the Lifshitz formula (see for example Ref.~\citen{Henkel04}).
In fact, in this limit, the Casimir energy 
is again completely dominated by the plasmonic 
contribution\cite{Van-Kampen68,Gerlach71,Henkel04}. 
Equation \eqref{smalldiss1} is valid 
also beyond the good conductor limit and could be used, e.g., to
analyze semiconductors where surface modes appear in a different
frequency range and can have much stronger damping~\cite{Joulain05}.

\subsection{Overdamped frequency modes}

As discussed in Sec.~\ref{sec:1-1} one of the consequences of the introduction of dissipation into the
system is the possible appearance of purely imaginary resonances corresponding to an overdamped behavior. As we will see below, they can also occur in Casimir physics and they play an important role in understanding a longstanding issue usually called plasma-Drude controversy~\cite{Bimonte07a,Torgerson04}. In simple terms, the controversy arises from experimental measurements of the Casimir interaction between two metallic objects (a sphere and a plane) which show a better agreement with the theoretical prediction of the plasma model than with that of the Drude model (see for example Ref.~\citen{Mostepanenko21} and references therein). This behavior is unexpected, since just on the physical basis of Ohm's law one should prefer Eq.~\eqref{DrudeModel} to Eq.~\eqref{plasmaModel}.

The quantum-thermodynamical properties of these overdamped frequency modes are to some extent unconventional and for our purposes it is convenient to analyze first the behavior of one of them.  The free energy of a mode with frequency $\Omega_{m}=-\imath\xi_{m}$ ($\xi$ positive) can be written as the sum of a ground state and a thermal contribution~\cite{Intravaia09,Intravaia10a,Reiche20}, i.e. $\mathcal{F}_{\xi}=\mathcal{E}^{0}_{\xi}+\Delta \mathcal{F}_{\xi}$, where~\cite{Nagaev02,Hanke95,Hanggi06,Hanggi08,Ingold09,Intravaia09,Henkel10}
\begin{equation}
\label{eq:E0}
	\mathcal{E}^{0}_{\xi}
		=
		 -\frac{\hbar\xi_{m}}{2\pi}\ln
		 \left[\xi\tau_{c}\right]>0
\end{equation}
and
\begin{equation}
\label{eq:FXi}
	\Delta\mathcal{F}_{\xi}
		=
		 \begin{cases}
			-\frac{\pi k_{\rm B}^{2}T^{2}}{6\hbar\xi_{m}}<0, 	& \hbar \xi_{m} \gg k_{\rm B}T \\
			\frac{k_{\rm B}T}{2}
				\ln[\frac{\hbar\xi_{m}}{k_{\rm B}T}] <0, 		& \hbar \xi_{m} \ll k_{\rm B}T
		\end{cases}.
\end{equation}
Notice that, differently from the ground state energy, the thermal contribution does not include a response time and features a different asymptotic behavior in the quantum ($\hbar \xi_{m} \gg k_{\rm B}T$) and classical limit ($\hbar \xi_{m} \ll k_{\rm B}T$).
It is interesting to notice here that the result for the classical limit can also occur if the resonance frequency of the modes depends on the temperature and vanishes faster than $T$ in the limit $T\to 0$. Curiously this implies that the system becomes more and more classical towards low temperature.

\begin{figure}
   \centering
   \includegraphics[width=\textwidth]{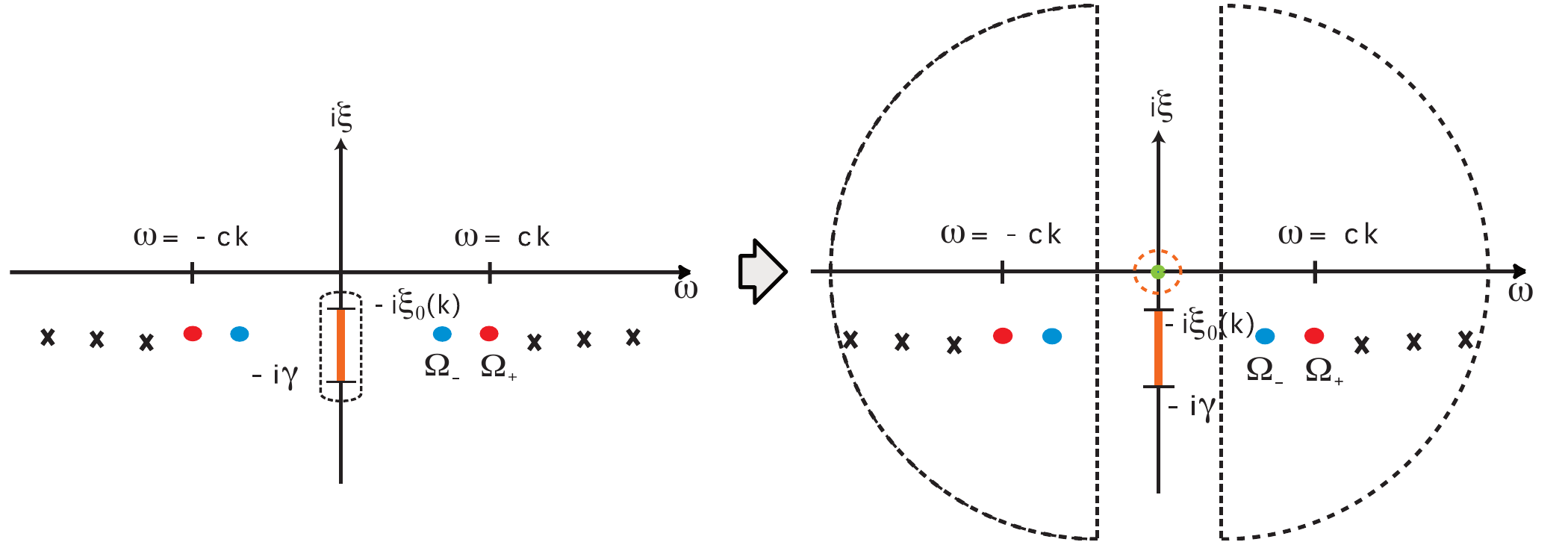} 
\caption{Equivalence of paths in the complex plane: (Left) Complex eigenfrequencies in the parallel plate geometry, for a fixed
value of $k$ (not to scale). 
Red and blue points: dissipative surface plasmons. 
Red line: continuum of eddy currents. Black crosses: propagating
modes.
(Right)
A counter-clockwise path around the eddy current continuum is equivalent 
to a clockwise path around the remaining part of the complex plane, encircling all other 
discontinuities. The total contour can be subdivided in the three contours appearing in the figure. The green dot describes the pole at $\zeta=0$ due to the expression of the free energy per mode [see Eq.~\eqref{eq:high-temp}]. Adapted from Ref.~\citen{Intravaia10}}
   \label{ChangePath}
   \label{BranchCut}
\end{figure}

With this information we can turn back to the Casimir effect for two identical semi-infinite
bulks facing each other in vacuum. If the material comprising the bulks is described by the Drude model, overdamped resonances appear as a continuum (the branch cut of the root 
$\kappa_{m}=\sqrt{k^{2}-\epsilon_{\rm D}(\omega)\omega^{2}/c^{2}}$) localized on the complex 
frequency plane along the negative imaginary axis. Specifically,
the cut is located between the branch points
$\zeta= -\imath\xi_{0}(\mathbf{k})
\approx -\imath \gamma(\lambda_{\rm p}/2\pi)^2 k^{2}$ (for $k \ll \omega_{\rm p} / c$)
and $\zeta = -\imath\gamma$ (see Fig.~\ref{BranchCut}). This means that all overdamped frequency modes are proportional to the dissipation rate $\gamma$~\cite{Intravaia09,Henkel10,Intravaia10a,Guerout14,Reiche20}.
From the physical point of view these resonances describe eddy currents circulating in the metal, i.e. low-frequency currents 
that satisfy a diffusion equation in the conductor~\cite{Jackson75} with a diffusion constant $D=\gamma(\lambda_{\rm p}/2\pi)^2$. These modes appear in addition to the dissipative 
generalization of the mode-frequencies considered in Sec.~\ref{nondissipativeModes}, showing that switching from the plasma to the Drude model is nontrivial and strongly alters the modes' structure of the system.
The contribution of the overdamped modes to the zero temperature Casimir energy can be defined from Eq.~\eqref{eq:Casimir-Diss} 
\begin{equation}
\label{eq:Eddy}
E_{\rm eddy} = \sum_{\sigma,\mathbf{k}} \,\left[\sum_{m}
-\frac{\hbar\xi^{\sigma}_{m}}{2\pi}
\ln[\xi^{\sigma}_{m}\tau_{c}]
\right]^{L}_{L\to\infty}= \int_{0}^{\infty}\!{\rm d}\xi
\sum_{\sigma,\mathbf{k}}\ 
\left[-
	\frac{\hbar\xi}{2\pi}\ln [\xi\tau_{c} 
\right]
\Delta^{\sigma}(\xi,L)~,
\end{equation}
where\footnote{Notice that for these modes alone, the sum rule in Eq.~\eqref{eq:Casimir-Diss} is not satisfied, indicating that their contribution to the Casimir energy depends on the response time $\tau_{c}$.} the last expression was obtained using the contour sketched in Fig.~\ref{BranchCut} (left),
and the differential density of overdamped modes given by
\begin{equation}
\Delta^{\sigma}(\xi,L)=\frac{1}{\pi}\partial_{\xi}\mathrm{Im} \ln \left[1-\{r^{\sigma}(-\imath \xi-0^{+},k)\}^{2}e^{-2\kappa L}\right].
\end{equation}
One can show~\cite{Intravaia09} that $E_{\rm eddy}$ gives rise to a repulsive contribution to the Casimir
force provided that $\gamma\tau_{c}\ll 1$. 
The contribution at high temperature to the free energy can be obtained from Eq.~\eqref{eq:Eddy} by replacing the minimal energy per overdamped mode with the classical limit $\mathcal{F}_{\xi}\sim k_{B}T \ln  (\hbar \xi / k_B T)$ [see Eq.~\eqref{eq:FXi}]. After a partial integration one can write~\cite{Intravaia09,Intravaia10a}
\begin{equation}
\label{eq:high-temp}
\mathcal{F}_{\rm eddy}\approx 
- \int_{0}^{\infty}\frac{d\xi}{\pi}\sum_{\sigma,\mathbf{k}}\ 
\frac{k_{B}T}{\xi} \mathrm{Im} \ln \left[1-\{r^{\sigma}(-\imath \xi-0^{+},k)\}^{2}e^{-2\kappa L}\right]~.
\end{equation}
The previous expression can be evaluated by noticing that the contour around
the eddy current continuum is equivalent to a contour encircling
the remaining part of complex plane~\cite{Intravaia10,Intravaia10a} and with it the poles of the integrand of Eq.~\eqref{eq:high-temp}, including all the other modes of the system
[Fig.~\ref{ChangePath} (right)]. 
The behavior of $\mathcal{F}_{\rm eddy}$ drastically differs as a function of the polarization also because of
the singularity $1/\xi$ explicitly appearing in the integrand of Eq.~\eqref{eq:high-temp}. For the TM-polarization
$\mathcal{F}^{\rm TM}_{\rm eddy}$ is small with respect to the contribution of the other (plasmonic and photonic) modes, i.e.
\begin{equation}
\mathcal{F}_{\rm Drude}^{\rm TM}=\mathcal{F}^{\rm TM}_{\rm eddy}+\mathcal{F}^{\rm TM}_{\rm pl}+\mathcal{F}^{\rm TM}_{\rm ph}\approx \mathcal{F}^{\rm TM}_{\rm pl}+\mathcal{F}^{\rm TM}_{\rm ph}.
\end{equation}
Physically, this can be understood by remembering that this polarization is dominantly electric near a material interface. 
The discontinuity of the electric field at the interface leads to charge accumulation that screens the interior of the bulk from electromagnetic fields, suppressing the interaction
with the slow-time dynamics occurring in its inside. 	This also explains why at short separations the total Casimir interaction
is well reproduced by the dissipative surface plasmon-polaritons discussed in the previous section, despite the fact that
the  eddy currents are associated with an evanescent field \cite{Intravaia09,Intravaia10a,Reiche20}. Since for good conductors the plasmonic and photonic modes are only slightly affected by dissipation one has
$\mathcal{F}_{\rm Drude}^{\rm TM}\approx \mathcal{F}_{\rm plasma}^{\rm TM}$~\cite{Klimchitskaya09c}.

The TE-polarization is particularly interesting because $r^{\rm TE}(\zeta \to 0) = 0$, indicating a lack of charge screening at the vacuum material interface.
This is consistent with the absence of surface plasmon-polaritons (see Fig.~\ref{disptetm}) and, in connection to the Bohr-van Leeuwen theorem~\cite{Leeuwen21}, with the mainly magnetic nature of 
the TE-polarized field.
Equation~\eqref{eq:high-temp} gives, up to a sign, the
same Casimir energy at high temperature (or large distance) of the photonic modes $\mathcal{F}^{\rm TE}_{\rm ph}$. Since these are only slightly affected by dissipation, they behave similarly in the Drude and plasma models. One then obtains
\begin{equation}
\label{eq:high-energy-diff}
\mathcal{F}^{\rm TE}_{\rm eddy} = -\mathcal{F}^{\rm TE}_{\rm ph}\approx -
\mathcal{F}^{\rm TE}_{\rm plasma} 
, \qquad (\gamma/\omega_{p}\ll 1)~.
\end{equation}
As a consequence, in the high-temperature limit $\mathcal{F}_{\rm Drude}^{\rm TE}=\mathcal{F}^{\rm TE}_{\rm eddy}+\mathcal{F}^{\rm TE}_{\rm ph} =0\neq\mathcal{F}^{\rm TE}_{\rm plasma}$. In other words, there is a compensation between eddy and photonic modes, leading to the disappearance of the TE-contribution~\cite{Intravaia09,Intravaia10,Intravaia10}, which is responsible for the difference in the theoretical prediction of the plasma and the Drude model. Specifically, this means that from the theoretical point of view the plasma-Drude controversy can be reframed as the need for a better understanding of the role of eddy currents in the Casimir effect.

The same analysis can be used for inspecting another thermodynamical aspect of the Casimir effect between two infinitely parallel bulks.
It was noticed~\cite{Bezerra04,Intravaia08,Klimchitskaya09b,Klimchitskaya09c,Milton11b} that if we consider an (infinite) bulk made of a metal with a perfectly periodically arranged background lattice (perfect crystal), intrinsic dissipation predominantly arises from scattering processes between the elementary particles in the system.  In this case dissipation is described by the Bloch-Gr\"{u}neisen formula which at low temperature predicts $\gamma(T)\propto T^m$ ($m\geq2$, $m=2$ for electron-electron scattering,  $m=3$ for s-d electron scattering and $m=5$ for electron-phonon scattering) \cite{Bloch29,Bloch30,Bass90}. 
Since all eddy currents' resonance-frequencies are proportional to the Drude dissipation rate, i.e. $\xi=\nu\;\gamma(T)$, they all feature the behavior $\hbar\xi/k_{\rm B}T\to0$ for vanishing temperature. 
As explained above, despite the low temperature limit, the eddy currents are behaving as if they were at high temperature, while all the other modes behave in a regular way, freezing to their ground state. 
This deeply affects the Casimir entropy, $\mathcal{S}=-\partial_{T}\mathcal{F}$. In the zero temperature limit the contribution to $\mathcal{S}$ stemming from the plasmonic and photonic modes goes to zero. Due to the charge screening affecting the TM-polarization, the entropy contribution due to the overdamped modes is dominated by the TE-polarization and from Eq.~\eqref{eq:high-energy-diff} we obtain
\begin{equation}
\mathcal{S}^{\rm TE}_{\rm eddy} (T\to 0)\approx -
\mathcal{S}^{\rm TE}_{\rm plasma}(T\to \infty)\equiv \mathcal{S}_{0}\neq 0
\end{equation}
It was pointed out that this behavior is contrary to what is usually expected from the third law of thermodynamics~\cite{Bezerra04,Intravaia08,Klimchitskaya09b,Klimchitskaya09c,Milton11b}, also known as the Nernst theorem, for which the entropy of a system must vanish at zero temperature. 
A similar anomaly was reported for the magnetic Casimir-Polder interaction between an atom and a metallic surface \cite{Reiche20,Haakh09b}.
The overdamped modes, to which the non vanishing entropy can be completely ascribed, provide a transparent description of the underlying physics.
Notice, however, that the previous result is very much related to the assumptions considered above, ranging from the perfect crystal limit and the absence of radiative damping, to the use of the Drude model. For example, it was reported that considering some form of residual dissipation, due for example to impurities~\cite{Bass90} ``regularizes'' the behavior of the entropy in the two plate configuration~\cite{Brevik05,Brevik06}. Similarly, without resorting to impurities, the result $\mathcal{S}(T\to 0)\to 0$ is obtained using the Lindhardt-Boltzmann-Mermin description~\cite{Svetovoy05,Svetovoy08a}, which considers a spatially nonlocal interaction between the material and the electromagnetic field and includes the Landau damping \cite{Landau46}. In this specific case a more in depth analysis \cite{Reiche20} has revealed that the ``regularization'' of the entropy behavior is imputable to the quantum properties introduced by the Fermi-Dirac statistics, characterizing the model. Physically, this is equivalent to a quantum-induced constraint on the diffusive dynamics of the eddy currents when the 
electrons' ballistic regime becomes dominant~\cite{Reiche20}.

\section{Nonequilibrium Casimir physics: the case of Quantum friction} 

Nonequilibrium systems play an important role in Casimir physics. Their understanding is important for a more accurate description of experimental setups and of configurations where unconventional effects are expected. 
Among the most common sources of nonequilibrium in Casimir physics one finds temperature gradients~\cite{Volokitin07,Biehs21}, external fields such as lasers~\cite{Bartolo16,Fuchs18a} and mechanical motion~\cite{Nation12,Reiche22}. Importantly, in all these cases an external agent that keeps the system from relaxing to equilibrium is involved.
The description of such systems is often
more complex than their equilibrium counterpart~\cite{Intravaia11} and has lead to the development of different approximation schemes. However, depending on the system these approximations must be used with care, since they can produce different scaling laws for the same phenomenon~\cite{Intravaia14,Intravaia16a} or simply fail in its quantitative description~\cite{Intravaia16,Intravaia19}.

Famous examples of nonequilibrium phenomena occurring when objects are set in motion are the Fulling–Davis–deWitt–Unruh effect and the dynamical Casimir effect~\cite{Nation12,Reiche22}. In both these examples one or more bodies violate the requirements for Lorentz invariance performing a non-inertial motion in vacuum. Due to this symmetry, despite the interaction with vacuum fluctuations, a motion at constant velocity in vacuum at zero temperature is preserved forever. The behavior changes, however, if it becomes possible to define a privileged frame with respect to which the dynamics at constant velocity occur. For example, if a neutral nonmagnetic particle is moving within a thermal field at $T\neq 0$, it will feel a frictional force hindering its motion. This phenomenon often goes under the name of black-body friction and can be related to the Einstein-Hopf drag~\cite{Einstein10,Milonni81,Ford85,Mkrtchian03,Lach12,Lach12a,Jentschura15,Oelschlager22,Guo21b}. Interestingly, a frictional force also appears when the motion at constant velocity is relative to one or more objects. Differently from black-body friction, however, the drag does not disappear in the limit $T=0$, highlighting in this specific case the role played by quantum fluctuations.

Although different setups have been considered~\cite{Pendry97,Volokitin11,Zhao12,Silveirinha14b}, this phenomenon, commonly known as quantum friction, is often investigated in the configuration involving an atom (or a microscopic object) moving parallel to a surface at constant height and velocity~\cite{Dedkov02a,Volokitin07,Intravaia15,Pieplow15,Scheel09,Farias20,Reiche22}.
In such a system the drag force can be roughly understood as arising from the interaction between the fluctuating dipole of the moving object and its image-dipole within the surface~\cite{Intravaia16a,Reiche22}. Due to dispersion and dissipation, the image is ``delayed" with respect to the real dipole, giving rise to a force having a component parallel to the surface. In general, if the stationary motion is non-relativistic and occurs with velocity $v$ parallel to translationally invariant objects, the quantum frictional force can be written as~\cite{Reiche20a,Intravaia19a,Reiche20c,Oelschlager22}
\begin{align}
  \label{Eq:ForceDipoleLate}
  F
  =
  -
  2
  \int_0^{\infty}\mathrm{d}\omega
  \int\frac{\mathrm{d}q}{2\pi}\;
  q\;
  \mathrm{Tr}
  \left[ 
  \underline{S}_{v}(-\omega_q^-)
  \underline{G}_{\Im}^{\mathsf{T}}(q,\mathbf{R}_a,\omega)
  \right]~.
\end{align}
In the previous expression $\underline{S}_{v}(\omega)$ is the velocity dependent atomic power spectrum tensor and $\underline{G}_\Im=(\underline{G}-\underline{G}^\dagger)/(2\imath)$, where $\underline{G}\equiv\underline{G}(q,\mathbf{R}_a,\omega)$ is the Green tensor describing the electromagnetic environment around the microscopic object at the position $\mathbf{R}_a$ in the plane orthogonal to the direction of motion. These quantities are evaluated at $q$, the component of the wave vector parallel to the axis of translational invariance, and $\omega_q^{\pm}=\omega\pm qv$, the Doppler-shifted frequency. The symbol $\mathrm{Tr}$ indicates the trace over the tensor product. Equation \eqref{Eq:ForceDipoleLate}, which does not depend on a specific model for the atomic system, describes the zero-temperature frictional force acting on the microscopic object when the system has reached its nonequilibrium steady state (NESS), i.e. when all transients have faded out.
The Green tensor in Eq.~\eqref{Eq:ForceDipoleLate} highlights the role of the material and the geometry of the bodies surrounding the moving object. In particular, one can show that for a planar structure the behavior of the quantum frictional interaction is mainly connected to the TM-polarized reflection coefficient through its imaginary part $\mathrm{Im}[r^{\rm TM}]\equiv r_{I}$. For a motion along the $x$-direction, the drag is dominated by the values $q=k_{x}\lesssim 1/z_{a}$ 
and frequencies $0<\omega\lesssim v/z_{a}$, where $z_{a}$ is the atom-surface separation.

Depending on the atom's velocity, quantum friction can feature a non-resonant and a resonant behavior~\cite{Volokitin07,Lach12,Lach12a,Jentschura15,Intravaia16,Intravaia16a}. The resonant behavior becomes relevant when the (mechanical) energy entering the system starts to match the energy of a system's resonance. 
Typically, this only occurs for very high velocities such that $v\gtrsim\omega_{\rm r}z_{a}$, where $\omega_{\rm r}$ is the resonance frequency under consideration (e.g. the atomic transition frequency or a surface polariton mode). 
The non-resonant interaction, which dominates at low velocities and therefore is more likely to occur in experiments,
 is usually connected with the dissipative tail of the lowest surface resonance and therefore it can be directly related to the low-frequency optical response of the material(s) comprising the substrate. For an atom moving above a homogeneous semi-infinite bulk made of an Ohmic material, like a metal with resistivity $\rho$ (e.g. for the Drude model $\rho=\gamma[\epsilon_{0}\omega_{\rm p}^{2}]^{-1}$), the non resonant frictional force takes the form \cite{Intravaia16}
\begin{eqnarray}
 F_\mathrm{bulk} \sim \Lambda\hbar\alpha_0^2\rho^2 \frac{v^3}{(2z_a)^{10} }
\,.
\label{eq:asyv}
\end{eqnarray}
where $\alpha_{0}$ is the atom's static polarizability. The numerical coefficient $\Lambda$ is specific to model for the atom~\cite{Intravaia16,Oelschlager18,Intravaia19}. The spatial dependency might also change, if one considers a spatially dispersive (nonlocal) material model~\cite{Reiche17,Reiche19,Volokitin02,Volokitin03a} or a nanoparticle instead of an atom~\cite{Intravaia14,Reiche20a}. In this last case, however, the force scales as the product of the dissipation rates of the materials comprising the particle and the surface instead of $\rho^{2}$~\cite{Intravaia14,Reiche20a}.

\begin{figure}
\centering
\parbox{5.5cm}{\includegraphics[width=5cm]{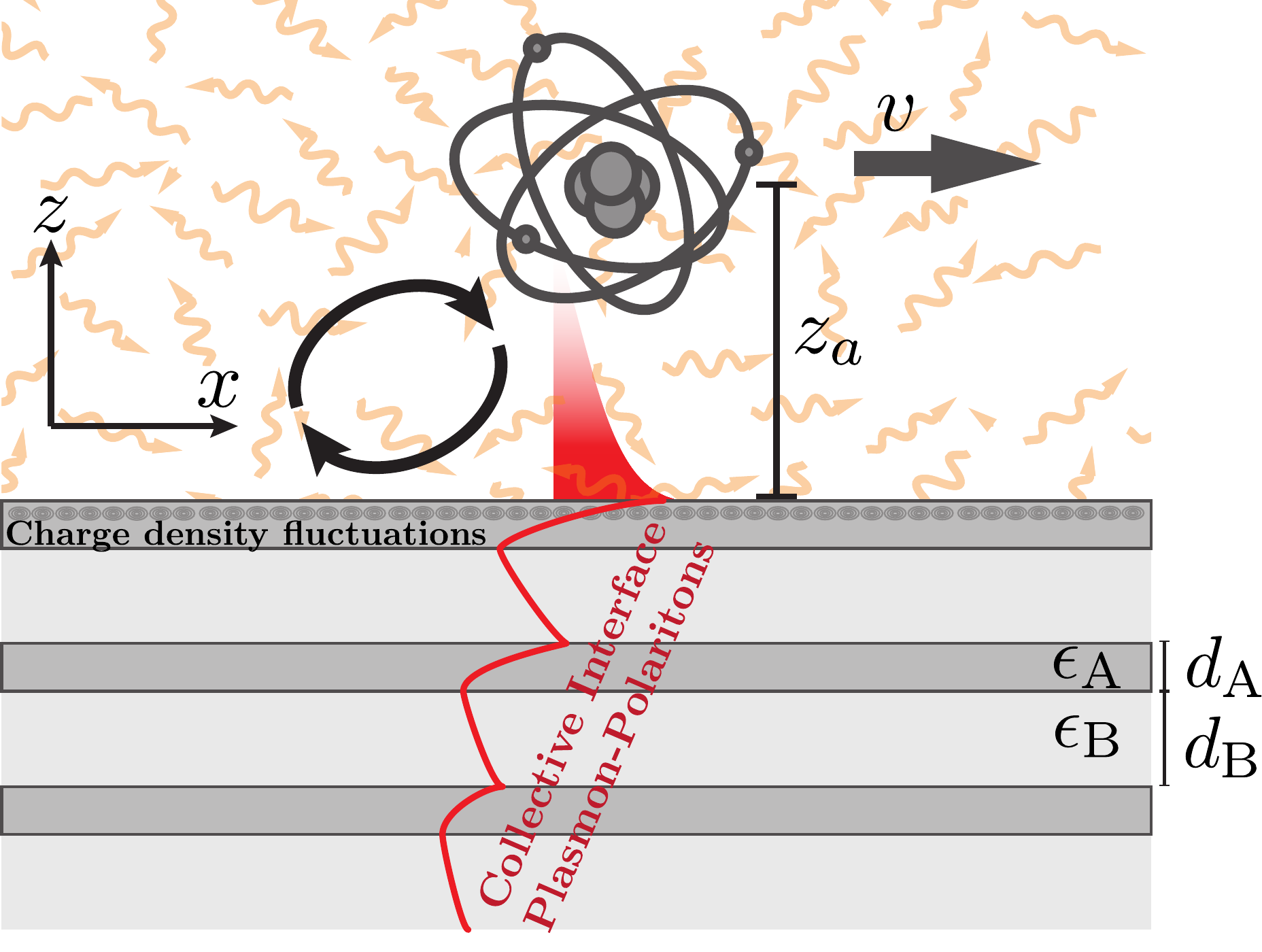}}
\parbox{6cm}{\caption{An atom moves at constant velocity and constant height above a half-space made by 
				a periodic sequence of alternating conductive ($\epsilon_{\rm A}$) and dielectric 
				 ($\epsilon_{\rm B}$) layers.
				 The spectrum of vacuum fluctuations is structured and the resulting quantum frictional force is sensitive to 
				 the interlayer interaction and the appearance of collective interface plasmon-polaritons (CIPP). Adapted from Ref.~\citen{Oelschlager18}.
\label{fig:setup}}}
\end{figure}

\begin{figure}
\resizebox{0.495\textwidth}{!}{
\includegraphics[scale=1]{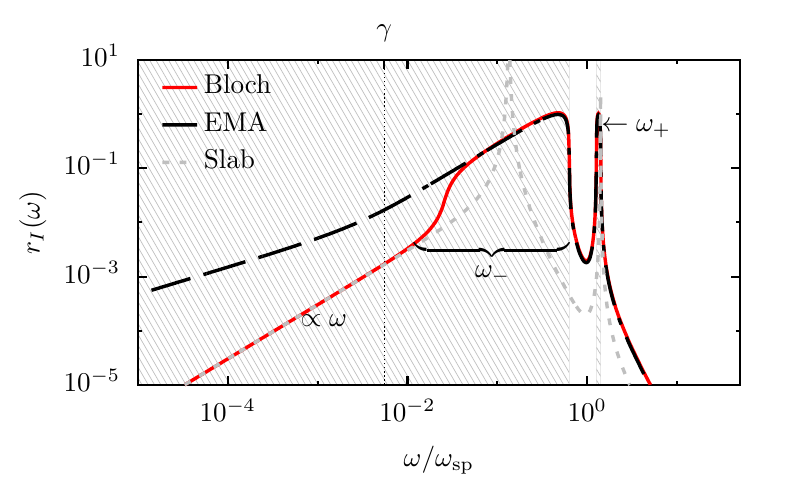}}
\resizebox{0.495\textwidth}{!}{
\includegraphics[scale=1]{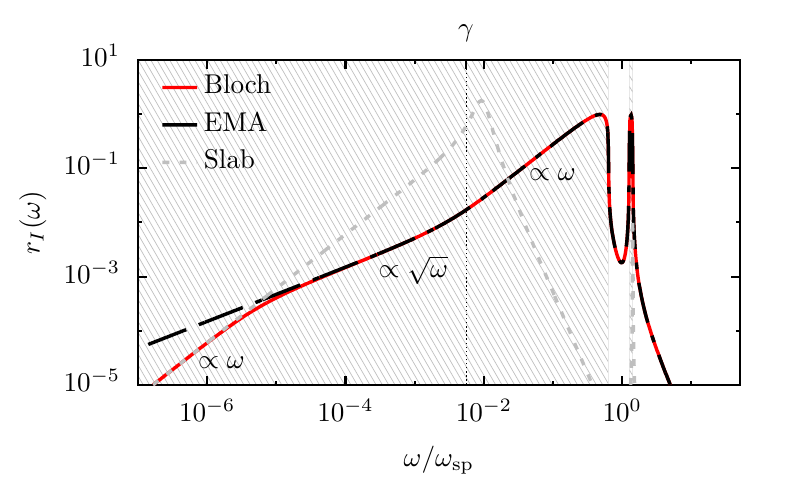}}
\caption{The imaginary part of the reflection coefficient (TM-polarization)  is 
				 plotted as a function of $\omega$ for different values of the in-plane wave vector. The conducting material's dielectric function, 
				 $\epsilon_{\rm A}(\omega)$, is described by the Drude model with parameters 
				 typical for gold \cite{Barchiesi14}. The dielectric is vacuum ($\epsilon_{\rm B}=1$),
				 the filling factor $f\equiv d_\mathrm{A}/(d_\mathrm{A}+d_\mathrm{B})=0.2$ and one has $k=10^{-1}c/\omega_\mathrm{p}$ in the left 
				 panel and $k=10^{-3}c/\omega_\mathrm{p}$ in the right panel.  Adapted from Ref.~\citen{Oelschlager18}.
				 \label{fig:refsqrt}
}
\end{figure}

One way to affect the behavior of the system's optical response and resonances is nanostructuring~\cite{Rodriguez13b}. For example, in multilayer structures made from alternating metallic and dielectric layers (see Fig.~\ref{fig:setup}), the surface modes, living at the dielectric-metal interface, start to interact across the structure, generating collective interface plasmon-polariton (CIPP) modes with different dispersion relations and shifted frequencies. Mathematically, the reflection coefficients of these multilayer structures can be exactly described using the transfer or the scattering matrix approach, often in combination with the Bloch theorem.
Figures~\ref{fig:refsqrt} depict the imaginary part of the (TM) reflection coefficient of the multilayer structure for two distinct values of the in-plane wave vector, in the case where the Drude model is used for describing the permittivity $\epsilon_{\rm A}(\omega)$ of the metallic layer while a constant, $\epsilon_{\rm B}$, is considered for the dielectric layer. When the periodic pattern is repeated a large (or also an infinite) number of times, the dispersion relations of the CIPP modes and consequently their resonance frequencies blur into a continuum that modifies the reflection coefficient as it is visible in Fig.~\ref{fig:refsqrt}, where the case of a single metallic slab is represented for comparison.
Three main regions appear~\cite{Camley84}: In analogy to the cavity result (see Sec.~\ref{nondissipativeModes}), two of these regions can be associated with symmetric ($\omega_-$) and antisymmetric ($\omega_+$) modes or equivalently their dissipative counterparts \cite{Johnson85}.
The third region can be better understood by first considering the so-called effective medium approximation (EMA)  \cite{Bruggeman35,Poddubny13}, which allows for a simple description of composite nanostructures such as those discussed above. The EMA relies on the fact that sufficiently large wavelengths cannot resolve the details of the multilayer system and consequently the nanostructuring fades in a homogeneous medium with an anisotropic permittivity $\underline{\epsilon}_{\rm EMA}(\omega) = \mathrm{diag}[\epsilon_\perp (\omega),\epsilon_\perp (\omega),\epsilon_\parallel(\omega)]$  (see e.g. Refs.~\citen{Mochan87,Poddubny13} for more details).
Interestingly, the EMA predicts that at a sufficiently low frequency, where the permittivities are such that $|\epsilon_{\rm A}|(\omega)\approx (\omega \epsilon_{0}\rho)^{-1}\gg \epsilon_{\rm B}$, the imaginary part of the reflection coefficient
behaves as 
\begin{equation}
  r_{I}(\omega,k)
	  \approx 
	\sqrt{\omega\epsilon_{0}} \sqrt{2\frac{\rho}{\epsilon_\mathrm{B}}\frac{d_{\rm B}}{d_{\rm A}}}~,
\label{rIsuplatt}
\end{equation}
featuring a behavior which is no longer Ohmic, i.e. linear in the frequency, but sub-Ohmic, i.e. proportional to the square root of the frequency. 
This behavior is visible in Figs.~\ref{fig:refsqrt}, where again the metal is described using the Drude model. 
In both plots, we can see that the EMA description enters the sub-Ohmic regime discussed above for 
$\omega<\gamma$ . This 
behavior is also featured by the calculation relying on the Bloch theorem (the transfer matrix approach), as long as the lower boundary of the $\omega_-$ branch lies below the value of the metal's damping rate $\gamma$. 
This occurs when, depending on the wave vector, the dispersion relation of the $\omega_-$ 
branch is stretched to lower frequencies. 
The sub-Ohmic trait of the superlattice occurs in the region where the modes of the 
$\omega_-$ branch become collectively overdamped~\cite{Oelschlager18}. 
Interestingly, both the Ohmic and the sub-Ohmic regions of the $\omega_-$ branch occur in a frequency range where the EMA predicts the appearance of the so-called hyperbolic range \cite{Poddubny13}.
In this region (shaded areas Figs.~\ref{fig:refsqrt}) a large number of wave vectors can be connected  with a narrow range of frequencies leading to a significant increase in the system's density of states \cite{Poddubny13}.
Figures~\ref{fig:refsqrt} also highlight the limit of the EMA approximation, underlining that an Ohmic behavior, corresponding to the first metal layer of the structure, is recovered when one considers sufficiently low frequencies.

Turning back to quantum friction, since the frequency range dominating the interaction increases with the velocity, the previous considerations indicate that, when the atom moves fast enough to start to ``perceive'' the substrate as being well-described by the EMA, the behavior of the force can significantly change with respect to the semi-infinite bulk case. 
One can show that such change of behavior is roughly expected when~\cite{Oelschlager18}
\begin{equation}
  vz_{a}\gtrsim \frac{d_{\rm A}d_{\rm B}}{2 \rho \epsilon_{0}\epsilon_{\rm B}}~.
\label{threshold}
\end{equation}
In this region the frictional force behaves quite differently from its low velocity counterpart 
leading to~\cite{Oelschlager18}
\begin{equation}
 F_\mathrm{EMA}\approx - \frac{6}{\pi^2}\hbar\alpha_0^2 \frac{\rho}{\epsilon_{0}\epsilon_\mathrm{B}}\frac{d_{\rm B}}{d_{\rm A}}\frac{v|v|}{(2z_a)^{9}}.
\label{velocity2}
\end{equation}
In comparison to the expression in Eq.~\eqref{eq:asyv} the frictional force no longer grows quadratically but linearly with the resistivity 
of the material. Remarkably, due to sub-Ohmic features, the force does not only change its velocity-dependence,  but also its functional behavior with respect to the atom-surface separation. 

\begin{figure}
\center
\resizebox{0.7\textwidth}{!}{
\includegraphics[scale=1]{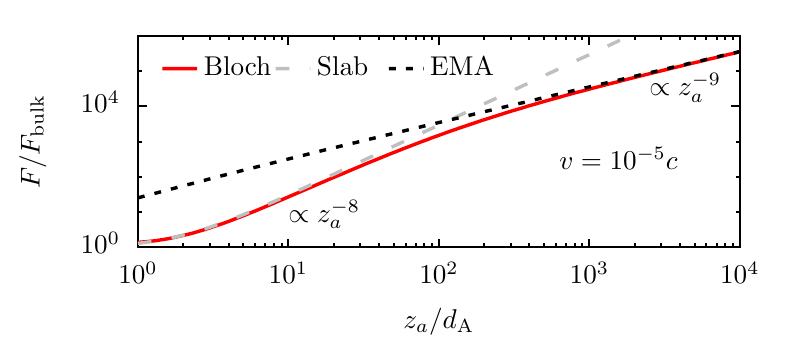}
}
\caption{Quantum friction as a function of the atom-surface separation. The force is normalized with respect to the semi-infinite bulk expression in Eq.~\eqref{eq:asyv} in order to highlight the corresponding enhancement of the interaction. When the atom is moving above a multilayer structure (red solid line) the frictional interaction becomes aware of the first layer, modifying its distance dependence from $z_{a}^{-10}$ to $z_{a}^{-8}$ (gray dashed line) before approaching the  EMA description (dotted black line), which yields a $z_a^{-9}$ law. The multilayer is made of alternating doped silicon-silicon layers. The first layer at the interface with vacuum is conducting. Adapted from Ref.~\citen{Oelschlager18}.
\label{fig:ztrans}
}
\center
\resizebox{0.532\textwidth}{!}{
\includegraphics[scale=1]{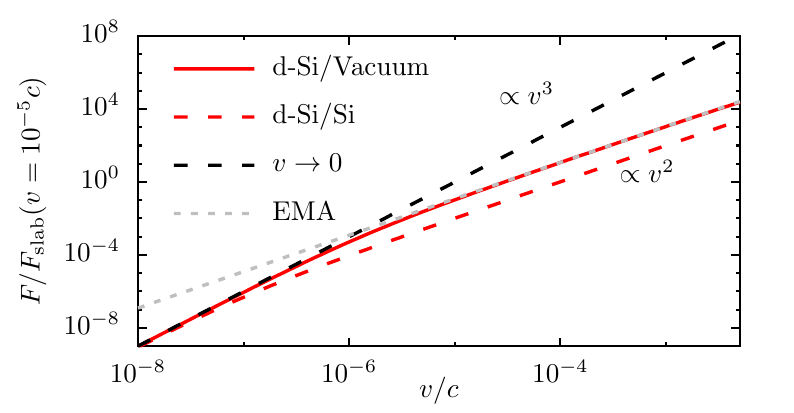}
}
\resizebox{0.456\textwidth}{!}{
\includegraphics[scale=1]{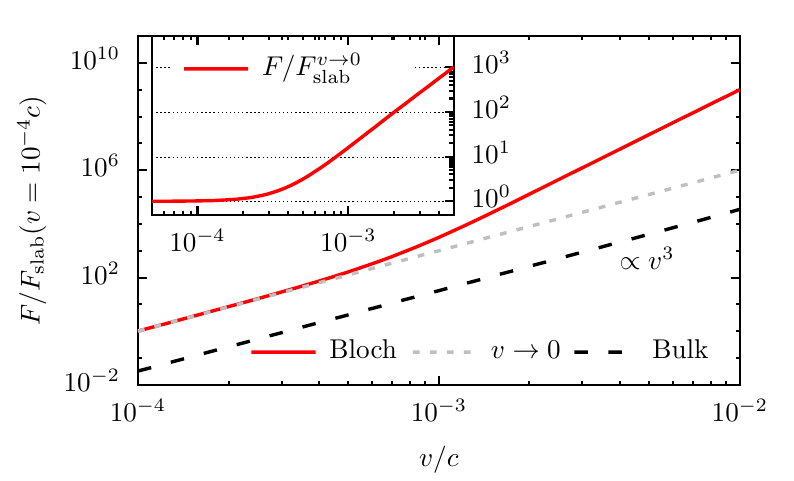}
}
\caption{(Left) Velocity dependence of the quantum frictional force for an atom moving at constant height above two different multilayer structures. 
In both cases, the transition from the $\propto v^{3}$ to  $\propto v^{2}$ behavior is visible. 
(Right) The resonant enhancement of quantum friction due to the collective interface plasmon-polariton (CIPP) modes within a multilayer structure (red solid line). 
A large atomic transition frequency, $\epsilon_{\rm B}=1$ and for the conductor a Drude metal with low damping constant were chosen in order to clearly reveal the effect of the CIPP modes. 
For comparison, the dashed black line describes the force with a Drude bulk substrate for which the resonant enhancement occurs at a larger velocity. 
				 \textit{Inset}: Quantum friction acting on an atom above a multilayer normalized by its low-velocity limit (grey dotted line in the main picture).  Adapted from Ref.~\citen{Oelschlager18}.
\label{fig:vtrans}}
\end{figure}

Figure \ref{fig:ztrans} depicts the quantum frictional force acting on an atom moving above a multilayer structure as a function of the atom-surface separation $z_{a}$ for $v/c=10^{-5}$. The force is normalized with respect to the expression in Eq.~\eqref{eq:asyv} in order to highlight the difference in the strength of the interaction and in its functional behavior.
We observe that the multilayers give rise to three different regimes. At short distances, 
we recover the bulk expression $ F\propto z_{a}^{-10}$ given in Eq.~\eqref{eq:asyv}. For
intermediate separations ($z_{a}\gtrsim d_{\rm A}$), the frictional interaction perceives only the first layer 
represented by a metallic slab. In this case the force behaves as $ F\propto 
z_{a}^{-8}$. Finally, for sufficiently large separations, the EMA regime is reached,
giving rise to a behavior $ F\propto z_{a}^{-9}$. 
The velocity dependence of the quantum frictional force is presented in Fig.~\ref{fig:vtrans}. 
For small velocities, where the Ohmic response of the structure dominates the interaction, the drag scales as $\propto v^3$ as in Eq.~\eqref{eq:asyv}. The region where the multilayer changes its behavior from Ohmic to sub-Ohmic becomes relevant at higher velocity [see Eq.~\eqref{threshold}], and the velocity dependence of the force changes from $\propto v^3$ to the $\propto v^2$, as described by Eq.~\eqref{velocity2}.

Finally, it is compelling to consider the resonant contribution to quantum friction in systems
in the presence of the previous multilayer structure. As discussed above, usually quantum friction's
resonant behavior occurs for rather high velocities, due to the fact that the relevant resonance
frequencies, such as those of surface plasmon-polaritons, $\omega_{\rm sp}$, are in the optical regime. In multilayer structures
the cross-interface coupling shifts these frequencies to below 
$\omega_{\rm sp}$, allowing for a certain degree of tunability via the thickness of the layers. 
This in turn lowers the velocity threshold for the occurrence of the resonant contribution to the
drag. The corresponding enhancement is visible in  Fig.~\ref{fig:vtrans}.
Physically, when the resonant condition is met, the continuum of modes in the $\omega_-$ branch effectively behaves as an efficient ``energy sink'' which tends to reduce the mechanical energy of the moving object transporting energy away from the surface through the multilayer structure.

\section{Conclusions}

Since the prediction of the Casimir effect in 1948, modes have played an important role in Casimir physics with more far-reaching implications than one could have initially expected.  Especially in quantum electrodynamics, frequency modes allow to formalize the connection between the zero-point energy and the existence of irreducible fluctuations in a system. These are two of the most classically unintuitive as well as fascinating aspects of the quantum theory and not rarely their investigation has revealed interesting and unexpected phenomena.

For example, we have seen that a modal analysis of the Lifshitz formula allows to reveal the contribution of two modes which have no counterpart in the evaluation of the Casimir force with perfect reflectors. These modes are connected with the existence of surface modes at the vacuum material interface. Although their relevance was already pointed out by van
Kampen to explain the van der Waals limit of the Lifshitz formula, further analysis has revealed that at larger separations they
provide a repulsive contribution which is balanced by the slightly larger attractive force arising from the remaining modes. This result has given rise to suggestions and experiments aiming to tailor the strength of the Casimir force by leveraging the properties of surface resonances.

Despite its utility, the definition of a mode becomes less transparent when dissipation exists in the system. Instead of modes one should speak of resonances or equivalently quasi-normal modes~\cite{Kristensen20}. Mathematically, the frequencies characterizing the modes become complex valued, rendering at least problematic the interpretation of the approach initially used by Casimir in 1948 but leaving unaffected the validity of the Lifshitz formula. The open quantum system paradigm provides the key to understand how dissipation modifies the expression for the ground state energy of a system and how the sum over mode approach needs to be modified to reestablish the equivalence with Lifshitz' result. Interestingly, also in this case a mode analysis reveals that what formally appears as a small modification of the Lifshitz formula has large repercussions on the mode spectrum. When dissipation is introduced, new (overdamped) quasi-normal modes appear in the system. They are characterized by a purely imaginary resonance frequency and may have curious thermodynamical properties. Their peculiarities help to understand the behavior of the Lifshitz formula in specific circumstances,

Finally, a modal analysis prouves very useful in characterizing and better understanding nonequilibrium phenomena. 
An example is quantum friction which describes the quantum mechanical drag felt by a particle moving with respect to one or more bodies. In particular, the quantum frictional force on a particle moving in vacuum ($T=0$) at velocity $v$ and at a height $z_{a}$ above a semi-infinite homogeneous bulk scales as $v^{3}/z_{a}^{10}$. This behavior changes, however, when the bulk is nanostructured: A change in the geometry and the material composition affects the bulk's optical response and induces a modification of the electromagnetic modes' spectrum. Specifically, when the bulk is replaced by a multilayer system alternating metallic and dielectric layers, the additional length-scales introduced in the system give rise to the appearance of a new sub-Ohmic regime, where the drag scales as $v^{2}/z_{a}^{9}$. This behavior can be associated with the existence of collective interface plasmon-polariton modes resulting from the interaction of the surface resonances existing at the metal-dielectric interfaces.

Given the importance of modes in Casimir physics, the material presented in this chapter is clearly not exhaustive. The few examples considered here should, however, highlight that understanding the often deeply intertwined connections between the mode structure of a system and phenomena typical of Casimir physics can lead to a better fundamental understanding of quantum mechanics and interesting experimental applications.

\section*{Acknowledgments}

The author thanks all his collaborators for enlightening discussions surrounding the topics reported in this chapter.
Special thanks go to Bettina Beverungen, Kurt Busch, Diego Dalvit, Carsten Henkel, Astrid Lambrecht, Marty Oelschl\"ager and Daniel Reiche for their work and support.

\bibliographystyle{ws-ijmpa}

\end{document}